\documentclass[a4paper, review]{cas-sc}
\usepackage[utf8x]{inputenc}

\usepackage[authoryear]{natbib}
\usepackage{caption}

\usepackage{subcaption}
\usepackage{lineno}
\usepackage{setspace}
\usepackage{aas_macros}
\usepackage[section]{placeins}

\begin{document}
\let\WriteBookmarks\relax
\def\floatpagepagefraction{1}
\def\textpagefraction{.001}
\shorttitle{The Excited Spin State of Dimorphos Resulting from the DART Impact}
\shortauthors{H.F. Agrusa et~al.}

\title[mode = title]{The Excited Spin State of Dimorphos Resulting from the DART Impact}

\author[1]{Harrison F. Agrusa}
\cormark[1]
\ead{hagrusa@astro.umd.edu}
\author[2]{Ioannis Gkolias}
\author[2]{Kleomenis Tsiganis}
\author[1]{Derek C. Richardson}
\author[3]{Alex J. Meyer}
\author[3]{Daniel J. Scheeres}
\author[4]{Matija \'{C}uk}
\author[5]{Seth A. Jacobson}
\author[6]{Patrick Michel}
\author[7]{\"{O}zg\"{u}r Karatekin}
\author[8]{Andrew F. Cheng}
\author[9]{Masatoshi Hirabayashi}
\author[6]{Yun Zhang}
\author[10]{Eugene G. Fahnestock}
\author[10]{Alex B. Davis}

\address[1]{Department of Astronomy, University of Maryland, College Park, MD, USA}
\address[2]{Department of Physics, Aristotle University of Thessaloniki, GR 54124 Thessaloniki, Greece}
\address[3]{Smead Department of Aerospace Engineering Sciences, University of Colorado Boulder, Boulder, CO, USA}
\address[4]{Carl Sagan Center, SETI Institute, Mountain View, CA, USA}
\address[5]{Department of Earth and Environmental Sciences, Michigan State University, East Lansing, MI, USA}
\address[6]{Universite C\^{o}te d'Azur, Observatoire de la C\^{o}te d'Azur, CNRS, Laboratoire Lagrange, Nice, France}
\address[7]{Royal Observatory of Belgium, Brussels, Belgium}
\address[8]{Johns Hopkins University Applied Physics Laboratory, Laurel, MD, USA}
\address[9]{Department of Aerospace Engineering, Auburn University, Auburn, AL, USA}
\address[10]{Jet Propulsion Laboratory, California Institute of Technology, Pasadena, CA, USA }
\cortext[cor1]{Corresponding author}

\begin{abstract}
The NASA Double Asteroid Redirection Test (DART) mission is a planetary defense-driven test of a kinetic impactor on Dimorphos, the satellite of the binary asteroid 65803 Didymos. DART will intercept Dimorphos at a relative speed of ${\sim}6.5 \text{ km s}^{-1}$, perturbing Dimorphos's orbital velocity and changing the binary orbital period. We present three independent methods (one analytic and two numerical) to investigate the post-impact attitude stability of Dimorphos as a function of its axial ratios, $a/b$ and $b/c$ ($a \ge b \ge c$), and the momentum transfer efficiency $\beta$. The first method uses a novel analytic approach in which we assume a circular orbit and a point-mass primary that identifies four fundamental frequencies of motion corresponding to the secondary's mean motion, libration, precession, and nutation frequencies. At resonance locations among these four frequencies, we find that attitude instabilities are possible. Using two independent numerical codes, we recover many of the resonances predicted by the analytic model and indeed show attitude instability. With one code, we use fast Lyapunov indicators to show that the secondary's attitude can evolve chaotically near the resonance locations. Then, using a high-fidelity numerical model, we find that Dimorphos enters a chaotic tumbling state near the resonance locations and is especially prone to unstable rotation about its long axis, which can be confirmed by ESA's Hera mission arriving at Didymos in late 2026. We also show that a fully coupled treatment of the spin and orbital evolution of both bodies is crucial to accurately model the long-term evolution of the secondary's spin state and libration amplitude. Finally, we discuss the implications of a post-impact tumbling or rolling state, including the possibility of terminating BYORP evolution if Dimorphos is no longer in synchronous rotation.
\end{abstract}

\begin{highlights}
\item High-fidelity numerical codes are essential for modeling the long-term spin evolution
\item DART may excite Dimorphos’ spin, leading to attitude instability and chaotic tumbling
\item Dimorphos is especially prone to unstable rotation about its long axis
\item A chaotic spin state will affect the system’s BYORP and tidal evolution
\item ESA's Hera mission may be able to place constraints on the system's tidal parameters
\end{highlights}

\begin{keywords}
Asteroids, dynamics \sep Celestial mechanics \sep Satellites of asteroids \sep Near-Earth objects \sep Rotational dynamics
\end{keywords}

\maketitle
\newpage
\doublespacing
\section{Introduction}

\begin{figure}
\centering
\includegraphics[trim={260 190 140 160},clip, width=1.0\textwidth]{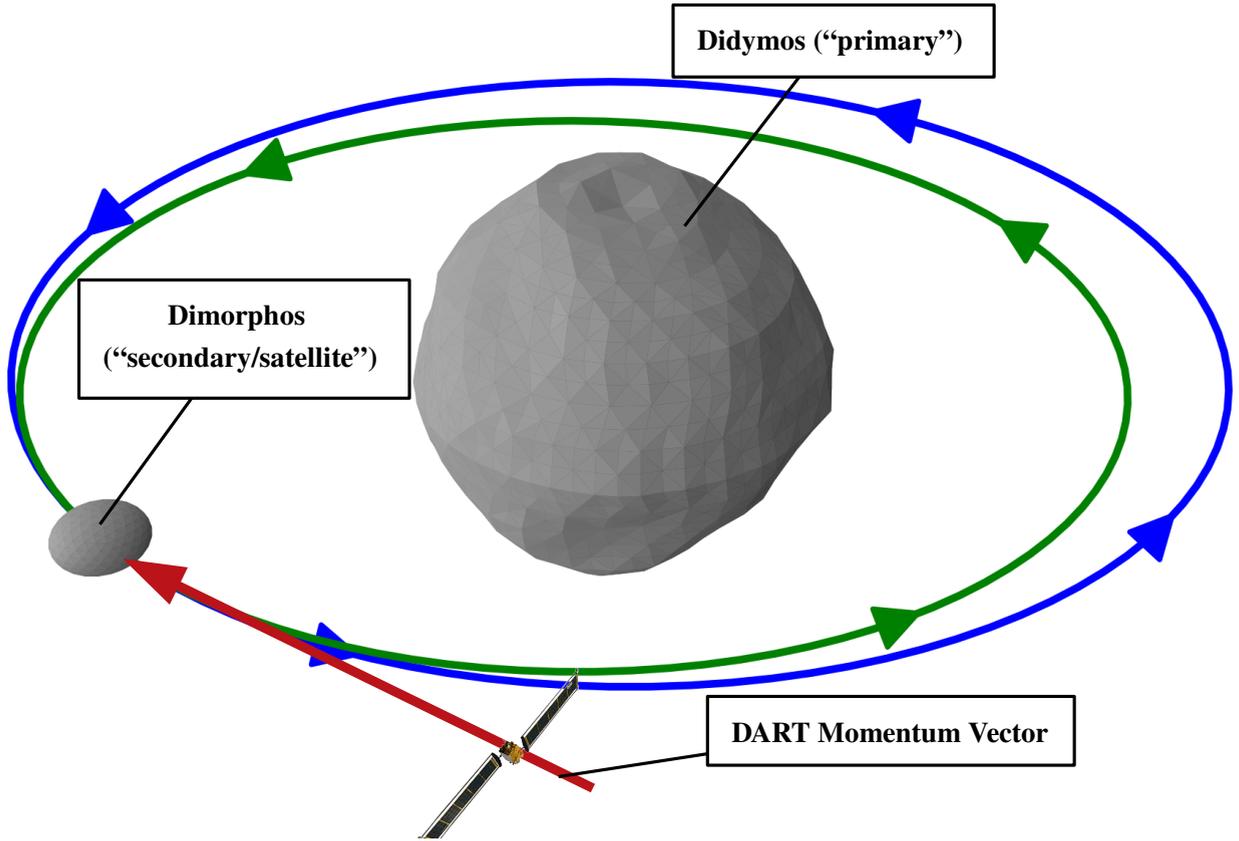}
\caption{\label{fig: didymosDiagram} Diagram showing the geometry of the problem and some terminology. In this work, ``Didymos'' or ``primary'' refers to the more massive, central component of the \textit{Didymos binary system}, while ``Dimorphos'', ``secondary'', or ``satellite'' refers to its smaller companion. The term ``impactor'' refers to the DART spacecraft. The red line denotes DART's momentum vector. In this work, this vector is assumed to lie within the plane of the mutual orbit, however the actual DART trajectory will have a nonzero inclination relative to the orbit plane. The blue line shows the shape and direction of the nominally assumed circular pre-impact orbit, while the green line shows an exaggerated post-impact orbit. This diagram shows the radar-derived polyhedral primary shape model from \cite{naidu2020} along with an assumed triaxial shape for the secondary. The body shapes and their mutual separation are to scale, while the DART spacecraft and post-impact orbit path are not. The spin poles of both bodies are assumed to be aligned with the mutual orbit pole.}
\end{figure}

NASA's Double Asteroid Redirection Test (DART) mission will be the first to demonstrate asteroid deflection by kinetic impact as a realistic assessment for planetary defense. The DART spacecraft will intercept the secondary (Dimorphos) of the near-Earth binary asteroid system 65803 Didymos in the fall of 2022 \citep{cheng2018}. The European Space Agency's (ESA) Hera mission will arrive at the binary ${\sim}4$ years later to investigate the resulting dynamical and geophysical changes to the system \citep{michel2018}. The nominal DART trajectory is an approximate head-on collision with Dimorphos, impulsively reducing its relative orbital speed, and thereby shortening the mutual orbit period and semimajor axis. The binary orbit eccentricity and inclination will also change, depending on the impact circumstances \citep{cheng2016}. Figure \ref{fig: didymosDiagram} shows a sketch of the binary system and the geometry of the problem. The change in orbit period will be measured with ground-based observations in order to infer  $\beta$, the momentum transfer efficiency. The change in velocity of an asteroid in response to a kinetic impact can be written as \citep{feldhacker2017, cheng2020},

\begin{equation}
\label{eq: deltaV_vector}
\Delta \vec{v} = \frac{m}{M}\bigg(\vec{u} + (\beta-1)(\hat{n}\cdot\vec{u})\hat{n}\bigg),
\end{equation}
where $m$ is the impactor mass, $M$ is the target mass, $\vec{u}$ is the impactor velocity, and $\hat{n}$ is the outward surface normal at the impact site. The first term represents the incident momentum of the spacecraft, and the second term is the contribution of escaping momentum, which is assumed to be along the surface normal. $\beta$ can then be written as the ratio of the total transferred momentum to the momentum delivered by the impactor:

\begin{equation}
    \beta = \frac{M(\hat{n}\cdot\Delta\vec{v})}{m(\hat{n}\cdot\vec{u})}.
\end{equation}
In reality, $\beta$ is a complicated function of the material properties and geometry of both the target and impactor \citep{stickle2020}. If we assume a head-on impact on a flat surface (allowing us to ignore the impact geometry), we can express $\beta$ as a simple function of scalars, 

\begin{equation}
  \beta = 1 + \frac{p_{\text{ejecta}}}{p_{DART}},
\end{equation}
where $p_{\text{DART}}$ is the scalar momentum carried by the DART spacecraft, and $p_{\text{ejecta}}$ is the scalar momentum carried by impact ejecta (which travels in the opposite direction). This expression for $\beta$ is much simpler than the equation used in practice, as it assumes that $p_{\text{ejecta}}$ and $p_{\text{DART}}$ are perfectly anti-aligned. However, this version is sufficient for describing \textit{why} $\beta$ is important: it tells us how much momentum is transferred to the target as a function of the impactor and ejecta momenta. For a more formal description and derivation of $\beta$, {\color{red} see \hypersetup{citecolor=red}\cite{rivkin2021}.}\hypersetup{citecolor=DarkSlateGrey} 

Due to the irregular shapes of both components and their close proximity, the spin and orbit of Dimorphos are highly coupled and non-Keplerian, meaning the dynamics cannot be treated as a simple point-mass 2-body problem. Therefore the use of high-fidelity, full-two-body-problem (F2BP) codes is crucial to understanding the complex dynamics \citep{agrusa2020}. Further, the shape of Dimorphos is still unknown and could have a major effect on the system's dynamics. With an assumed triaxial ellipsoid shape for Dimorphos, we explore the post-impact dynamical evolution of the system as a function of the possible axial ratios of the secondary and the momentum transferred by the DART impact ($\beta$). In Section \ref{subsection: background}, we give some brief background on the Didymos binary and the DART impact's implications for the secondary's libration state. Then Section \ref{section: methods} introduces our novel analytic approach and two numerical methods for studying the spin dynamics of Dimorphos. The results for each of these three methods are presented in Section \ref{section: results}. Finally, we discuss the implications of our results in Section \ref{section: discussion}.
\subsection{Background}
\label{subsection: background}
Although it has not yet been confirmed with observations we nominally assume that Dimorphos is in the 1:1 spin-orbit resonance (i.e., tidally locked)\footnote{\cite{naidu2020} find that the radar bandwidth of secondary is consistent with a spin period equal to the orbit period, suggesting Dimorphos may be in synchronous rotation.}. Didymos's spinning-top shape and fast rotation are suggestive of a rubble-pile structure, owing to likely formation scenarios such as spin-up-driven mass loss (followed by gravitational accumulation of the secondary), or gravitational reaccumulation after a catastrophic disruption \citep{richardson2006}. In addition, its spin rate exceeds the spin barrier at the nominal bulk density of ${\sim}2.17 \text{ g cm}^{-3}$, implying some level of interparticle cohesion and/or higher bulk density \citep{zhang2017, zhang2018, zhang2021}. If Dimorphos and Didymos have a common origin, this suggests that Dimorphos is also a rubble pile. The highly dissipative nature of rubble-pile asteroids implies that the system has had sufficient time for Dimorphos to become tidally locked and enter a dynamically relaxed state \citep{goldreich2009, jacobson2011a}. For these reasons we assume the system's pre-impact dynamical state is relaxed, meaning the mutual orbit is well-circularized with the secondary in the 1:1 spin-orbit resonance and any free libration is minimized. However, it should be noted that observations have not confirmed such a relaxed state, rather it just has not been ruled out \citep{pravec2006, scheirich2009, naidu2020}. If, upon arrival at the Didymos system, we find that the mutual orbit and secondary spin are already excited, the DART impact will likely further excite the mutual dynamics. Therefore, the results presented in this work should be interpreted as a conservative estimate of the possible impact outcomes.

\subsubsection{Libration Concepts}
The angle between the line-of-centers (LOC) and the secondary's long axis is commonly referred to as a \textit{libration} angle. In the classic (uncoupled) spin-orbit problem, there are two distinct libration modes: \textit{free} and \textit{forced} \citep{MurrayDermott, naidu2015}. Although this paper explores the dynamics of the fully coupled spin and orbital dynamics of the Didymos-Dimorphos system, the insights from the classic spin-orbit problem provide useful intuition for understanding the dynamics when we consider the full problem. For a circular, uncoupled planar orbit, a first-order approximation for the frequency of free libration is given by (see Ch. 5 of \cite{MurrayDermott}),
\begin{equation}
\label{eq: librationFrequency}
  \omega_{\text{lib}} = n\bigg(\frac{3(B-A)}{C}\bigg)^{1/2},
\end{equation}
where $n$ is the mean motion, and $A$, $B$, and $C$ are the secondary's three principal moments of inertia (which correspond to the axis lengths $a \ge b \ge c$). For certain combinations of the three moments of inertia, the free libration frequency can become resonant with the forced libration frequency (i.e., the mean motion) and a \textit{secondary} resonance can occur \citep{Melnikov2001, gkolias2019}. This can lead to an intricate dynamical environment, which only becomes more complicated when we allow for non-zero eccentricity, out-of-plane motion, and a full coupling between the mutual orbit and the spin states of both bodies. 

It is important to note that the DART impact will excite both free and forced libration modes, even if they have been damped to a minimum prior to the impact. The velocity perturbation from DART will increase the binary eccentricity \citep{cheng2016}, increasing the \textit{forced} libration mode, due to the restoring torque that the secondary feels as it becomes misaligned with the LOC as the orbital angular velocity changes throughout the orbit. With a nearly instantaneous perturbation to the orbital velocity of the secondary, DART will also induce \textit{free} libration modes by creating a difference in its instantaneous orbital and spin angular velocities. 

In reality, Dimorphos's attitude has three degrees of freedom relative to the uniformly rotating orbit frame (roll, pitch, and yaw) and the system could have a nonzero eccentricity and inclination. Therefore, its spin evolution will be more complicated than the two idealized libration modes used here as a conceptual example. Namely, the excited planar libration modes, for particular shapes of the secondary, can induce significant out-of-plane rotation \citep{Kane1965, Eapen2021}. Moreover, energy transitions can happen between the planar and out-of-plane rotational degrees of freedom that is attributed to resonant phenomena \citep{breakwell1965}. We will see that the excitation of Dimorphos's libration state, primarily due to the excitation of nonplanar rotation, can lead to chaotic motion. Chaotic rotation has been observed for many other bodies in our solar system such as the triple system (47171) Lempo, Saturn's Hyperion, and Pluto's outer four satellites, to name a few examples \citep{correia2018, Wisdom1984, showalter2015}. 

\subsubsection{Euler Angles}
In this work, we treat the ``libration angle'' as simply the angle between the long-axis of the secondary and the line-of-centers. As described above, in the classic spin-orbit problem, this angle would be purely within the plane of the orbit. However, we will see that this angle will have nonplanar components if the secondary's attitude becomes unstable.

\begin{figure}
\centering
\includegraphics[trim={10 5 0 0},clip, width=0.4\textwidth]{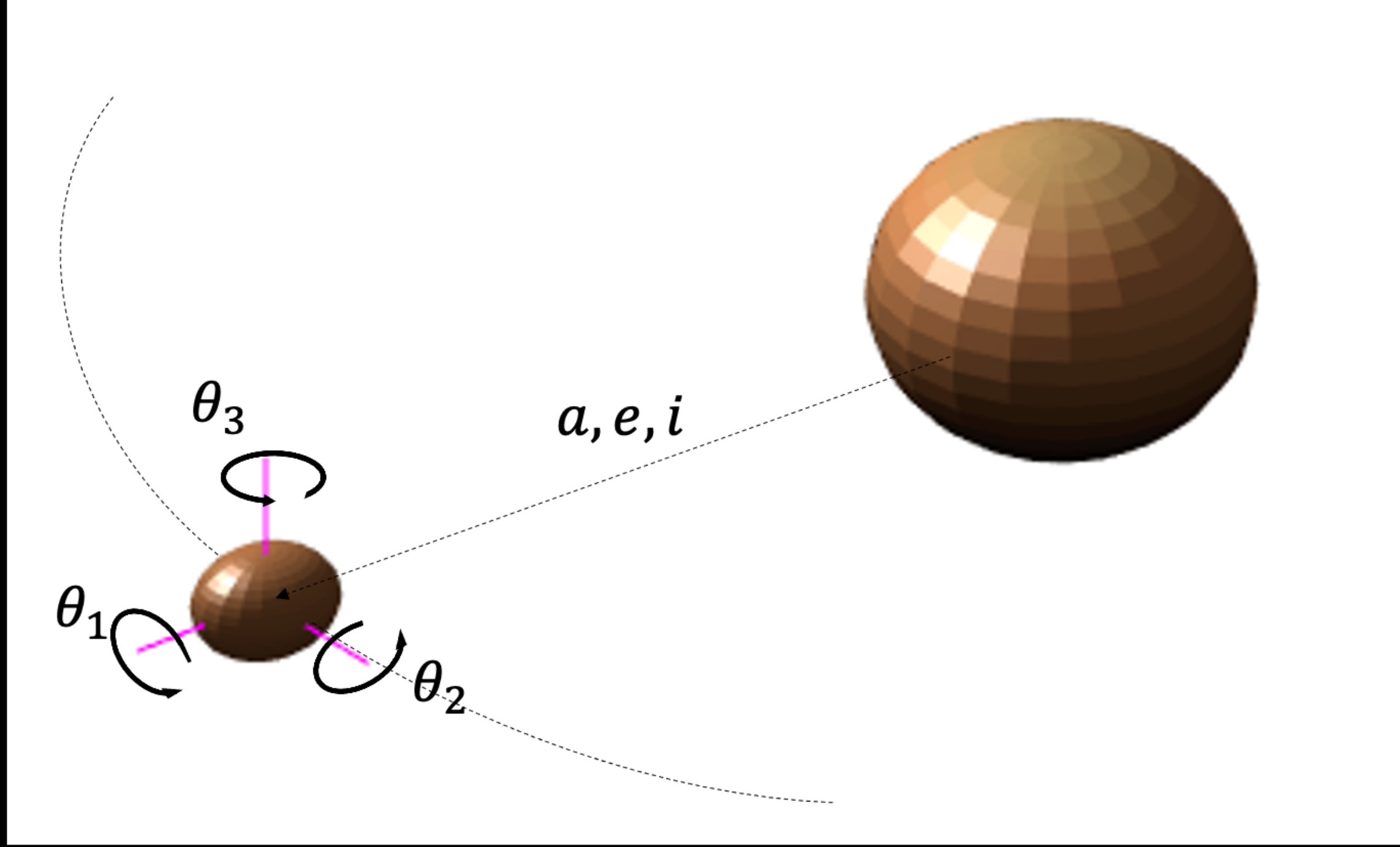}
\caption{\label{fig: eulerIllustration} Schematic showing the Euler angle set for Dimorphos. $\theta_{1}$ is the roll angle, $\theta_{2}$ is pitch, and $\theta_{3}$ is yaw. In the diagram, the pink axis markers point \textit{out} of Dimorphos, and the angles rotate in the sense of the right-hand rule. The angles are defined such that all three angles are equal to zero when the secondary's body-fixed frame is aligned with the orbital frame. (Image credit: \cite{meyer2021})}
\end{figure}

Instead of just looking at the libration angle, we can examine all three Euler angles that make up the secondary's attitude. We use the 1-2-3 Euler angle set (roll-pitch-yaw) shown in the diagram on Fig.\ \ref{fig: eulerIllustration}, where the Euler angles give Dimorphos's attitude in the frame rotating with the orbit. At each simulation output the rotating frame is defined as follows: the $x$-axis points along the LOC, the $z$-axis is the direction of the mutual orbit pole (i.e., the orbital angular momentum vector), and the $y$-axis completes the right-handed triad. A direction cosine matrix between the secondary's body-fixed frame to the rotating frame is constructed, from which the three Euler angles are computed. See Appendix B of \cite{schaub2009} for the precise mathematical derivation of this Euler angle set.

The Euler angles $\theta_{2}$ (pitch) and $\theta_{3}$ (yaw) can be thought of as two libration angles; $\theta_{2}$ is analogous to an out-of-plane (latitude) libration, and $\theta_{3}$ is analogous to the planar (longitude) libration. Due to the ordering sequence of the Euler angles, this is only technically true when $\theta_{1}$ is exactly zero, which we will see is not the case. However, thinking about $\theta_{3}$ and $\theta_{2}$ as the respective planar and non-planar components of the secondary's libration can be a useful conceptual tool.

\section{Methods}
\label{section: methods}

In Section \ref{subsection: linearizedDynamics}, we use an analytic approach to investigate the attitude stability and presence of resonant libration frequencies of Dimorphos, under the assumption that the primary is a uniform sphere and that the system is in an equilibrium state (i.e., a circular orbit)\footnote{The assumption that the pre-impact orbit is relaxed and circular is made for a few reasons. First, ground-based observations have not placed a lower limit on the mutual eccentricity and three studies have derived upper limits of 0.03, 0.04, and 0.05 (\cite{scheirich2009}, \cite{fang2012}, \cite{naidu2020}, respectively). Second, if the two components are rubble piles (for reasons laid out in Section \ref{subsection: background}), then we would expect tides to quickly damp the system to an equilibrium state \citep{goldreich2009}. Third, assuming a relaxed pre-impact state means that the post-impact dynamics predicted by this work can be interpreted as a conservative estimate. In other words, the predictions made in this work should be considered a ``lower limit'' for the excitation of Dimorphos's spin state resulting from the DART impact. If the binary is already excited prior to DART, then the impact may only further excite the system.}. Then we employ two numerical codes to study the attitude dynamics of Dimorphos following a DART-like perturbation to the relative velocity of Dimorphos. The first code, described in Section \ref{subsection: simplified3Dmomdel}, computes a coupled planar (2D) orbit between an oblate spheroid (Didymos) and a triaxial ellipsoid (Dimorphos) and parametrically inserts the solution into Euler's rigid body equations to follow Dimorphos's attitude in 3D. In Section \ref{subsection: GUBAS}, we introduce the second numerical code, which uses the radar-derived shape model for Didymos and computes the fully coupled spin and orbital evolution of the system. To clarify, all three models used in this work have at least some degree of spin-orbit coupling, each with increasing levels sophistication. Finally, our approach to setting up the initial conditions for the numerical simulations is detailed in Section \ref{subsection: probSetUp}. 

\subsection{Analytic Dynamics Model}
\label{subsection: linearizedDynamics}
Here we develop an analytic, linearized approach to calculating Dimorphos's fundamental frequencies as a function of its semi-axes. Previous work has addressed this problem using a variety of simplifying assumptions. The first to estimate the uncoupled frequencies was Lagrange in his 1764 essays on the libration of the Moon \citep{fraser1983}. Modern day derivations can be found in various textbooks; \cite{MurrayDermott} provide the planar libration frequency (i.e., Eq. \eqref{eq: librationFrequency}) and \cite{curtis} gives the non-planar frequencies. Also, \cite{Fleig1970} derives fundamental frequencies for a triaxial satellite, with a good discussion on the role of resonances with the offplane frequencies (see Appendix \ref{appendix: natural frequencies}). \cite{scheeres2006} developed an approach for this computation using an arbitrary body and a sphere, which is used by \cite{fahnestock2008simulation} to solve for analytic expressions of the four fundamental frequencies around a synchronous equilibrium applied to the binary system Moshup (previously 1999 KW4). In the simplest analytic approximation, \cite{fahnestock2008simulation} assume a spherical primary and report the frequencies as a function of the secondary's shape. This is further expanded to fourth-order by \cite{boue2009spin}, where the mutual interactions between the bodies are considered and report good estimates for the precession and nutation frequencies for Moshup. For our purposes, the simpler approach outlined in \cite{scheeres2006} and \cite{fahnestock2008simulation} is sufficient to provide an idea of the expected behavior in the system, which we expand upon here.

The potential energy between two bodies taken to a second-order approximation is \citep{scheeres2009}:
\begin{equation}
V = \frac{-GM_1M_2}{R}-\frac{G}{2R^3}[M_1\text{Tr}(I_2)+M_2\text{Tr}(I_1)]+\frac{3G}{2R^5}\vec{R} \cdot [M_1A_2^TI_2A_2+M_2A_1^TI_1A_1] \cdot \vec{R}
\end{equation}
where $A_i$ is the matrix transforming from the frame in which the relative position vector $\vec{R}$ is specified to body-fixed coordinates, with the subscript 1 and 2 referring to the primary and secondary, respectively. The inertia tensors of the primary ($I_{1}$) and secondary ($I_{2}$), are written in their respective body-fixed frames. For the spherical restricted full two-body problem, body 1 is a sphere while body 2 is an arbitrary 3D massive asteroid. Since body 1 is a sphere, the term $\vec{R}\cdot A_1^TI_1A_1 \cdot \vec{R}$ can simply be written as $R^2I_S$ where $I_S$ is the diagonal entry of $I_1$ (i.e., the moment of inertia of a uniform sphere) and $\text{Tr}(I_1)$ reduces to $3I_S$, thus the higher-order terms involving body 1 disappear.

If the position vector $\vec{R}$ is written in the frame of body 2, $A_2$ then becomes the identity matrix. With these simplifications, the second-order potential energy becomes:
\begin{equation}
V = \frac{-GM_1M_2}{R}-\frac{G}{2R^3}[M_1\text{Tr}(I_2)]+\frac{3G}{2R^5}\vec{R} \cdot [M_1I_2]\cdot \vec{R}.
\end{equation}

The equations of motion for this model, taken in the body-fixed frame of the secondary, are:
\begin{align}
\ddot{\vec{R}}+2\vec{\Omega}\times\dot{\vec{R}}+\dot{\vec{\Omega}}\times\vec{R}+\vec{\Omega}\times(\vec{\Omega}\times\vec{R})=-\frac{1}{m}\frac{\partial V}{\partial \vec{R}},\\
I\dot{\vec{\Omega}}+\vec{\Omega}\times I\Omega = \vec{R}\times\frac{\partial V}{\partial \vec{R}},
\end{align}
where we drop the subscript on $I$, which is the inertia tensor of the secondary (body 2). Here, $\vec{\Omega}$ is the angular velocity of the frame of the secondary and $m$ is the reduced mass: $m=\frac{M_1M_2}{M_1+M_2}$. 

This problem can be normalized by introducing a length scale $\alpha$ (equal to the body separation) and mean motion $n = \sqrt{G(M_1+M_2)/\alpha^3}$. With this convention, we define,
\begin{align}
r &\equiv \frac{R}{\alpha} \\  \omega &\equiv \frac{\Omega}{n} \\  \mathcal{I} &\equiv \frac{I}{M_2\alpha^2}.
\end{align}
The equations of motion can now be rewritten as,
\begin{equation}
\ddot{\vec{r}}+2\vec{\omega}\times\dot{\vec{r}}+\dot{\vec{\omega}}\times\vec{r}+\vec{\omega}\times(\vec{\omega}\times\vec{r})=-\frac{\partial \mathcal{V}}{\partial \vec{r}},
\end{equation}
\begin{equation}
\mathcal{I}\dot{\vec{\omega}}+\vec{\omega}\times \mathcal{I}\omega = \nu\vec{r}\times\frac{\partial \mathcal{V}}{\partial \vec{r}},
\end{equation}
where we introduce a mass fraction $\nu=\frac{M_1}{M_1+M_2}$ ($\nu \simeq 0.99$ for the Didymos system) and the normalized potential energy,

\begin{equation}
\mathcal{V} = \frac{-1}{r}-\frac{1}{2r^3}[\text{Tr}(\mathcal{I})]+\frac{3}{2r^5}\vec{r} \cdot [\mathcal{I}]\cdot \vec{r}.
\end{equation}

Defining the state vector as $\vec{X} = [\vec{r},  \dot{\vec{r}}, \vec{\omega}]^T$, the state dynamics can thus be written as,
\begin{equation}
\dot{\vec{X}} = 
\begin{bmatrix} \dot{\vec{r}} \\
			-2\tilde{\vec{\omega}}\dot{\vec{r}} + \tilde{\vec{r}}\dot{\vec{\omega}} - \tilde{\vec{\omega}}\tilde{\vec{\omega}}\vec{r} -\frac{\partial \mathcal{V}}{\partial \vec{r}} \\
			\mathcal{I}^{-1}\left[-\tilde{\vec{\omega}}\mathcal{I}\vec{\omega} + \nu\vec{r}\times\frac{\partial \mathcal{V}}{\partial \vec{r}}\right]
			\end{bmatrix}
			= \vec{F}(\vec{X}),
\end{equation}
where we introduce the tilde notation for the cross product skew-symmetric operator. Then, the equilibrium conditions are:
\begin{equation}
\dot{\vec{r}}=\dot{\vec{\omega}}=0
\end{equation}
\begin{equation}
\tilde{\vec{\omega}}\tilde{\vec{\omega}}\vec{r} = -\frac{\partial \mathcal{V}}{\partial \vec{r}}
\end{equation}
\begin{equation}
\tilde{\vec{\omega}}\mathcal{I}\vec{\omega}=\nu\vec{r}\times\frac{\partial \mathcal{V}}{\partial \vec{r}}.
\end{equation}

With these equilibrium conditions, the linearized dynamics matrix at equilibrium can be calculated:
\begin{equation}
\frac{\partial \vec{F}}{\partial \vec{X}} \biggr\rvert_{\vec{X}_0} = 
\left[
\arraycolsep=1.4pt\def\arraystretch{2.2}
\begin{array}{c|c|c}
[\bold{0}]_{3\times3} & [\bold{U}]_{3\times3} & [\bold{0}]_{3\times3} \\ \hline
\nu\tilde{\vec{r}}\mathcal{I}^{-1}\left[\tilde{\vec{r}}\frac{\partial^2\mathcal{V}}{\partial \vec{r}^2}-\widetilde{\frac{\partial \mathcal{V}}{\partial \vec{r}}}\right] - \tilde{\vec{\omega}}\tilde{\vec{\omega}} - \frac{\partial^2\mathcal{V}}{\partial \vec{r}^2} & 
-2\tilde{\vec{\omega}} & 
\tilde{\vec{r}}\mathcal{I}^{-1}\left[-\tilde{\vec{\omega}}\mathcal{I} + \widetilde{\mathcal{I}\vec{\omega}}\right] + \tilde{\vec{\omega}}\tilde{\vec{r}} + \widetilde{\tilde{\vec{\omega}}\vec{r}} \\ \hline
\nu\mathcal{I}^{-1}\left[-\widetilde{\frac{\partial \mathcal{V}}{\partial \vec{r}}}+\tilde{\vec{r}}\frac{\partial^2\mathcal{V}}{\partial \vec{r}^2}\right] &
[\bold{0}]_{3\times3} & \mathcal{I}^{-1}\left[-\tilde{\vec{\omega}}\mathcal{I} + \widetilde{\mathcal{I}\vec{\omega}}\right]
\end{array}
 \right],
\end{equation}
where $ [\bold{U}]_{3\times3}$ is the unitary matrix. This gives the linearized dynamics equation about the equilibrium:
\begin{equation}
\begin{bmatrix} \delta\dot{\vec{r}} \\ \delta\ddot{\vec{r}} \\ \delta\dot{\vec{\omega}} \end{bmatrix}
= \frac{\partial \vec{F}}{\partial \vec{X}} \biggr\rvert_{\vec{X}_0}
\begin{bmatrix} \delta\vec{r} \\ \delta\dot{\vec{r}} \\ \delta\vec{\omega} \end{bmatrix}.
\end{equation}

\cite{fahnestock2008simulation} calculate the dynamics matrix to obtain a simplified expression, from which they derive equations for the four fundamental frequencies after reducing the matrix to $8\times8$. We will step through the process of reducing this matrix using the angular momentum integral. However, rather than obtaining expressions for the frequencies, we will directly solve for them using spectral decomposition of the reduced dynamics matrix.

The magnitude of the angular momentum provides an integral of motion allowing us to reduce the dynamics matrix from $9\times9$ to $8\times8$. In practice, this leads to eliminating the $\omega_3$ ($x_9$) contribution, which would otherwise result in a zero eigenvalue. To reduce the matrix, we can break the problem up as:

\begin{equation}
\delta\dot{\vec{X}} = 
\begin{bmatrix} A_y & A_{8\times1} \\ A_{1\times8} & A_9 \end{bmatrix}
\begin{bmatrix} \delta\vec{y} \\ \delta x_9 \end{bmatrix}
\end{equation}
where $\vec{X}$ is the state, $\vec{y}$ is the first 8 states (excluding $x_9$), and the full linearized dynamics matrix is called $A$, which we have broken up into convenient submatrices. This allows us to write

\begin{equation}
\delta{\dot{\vec{y}}} = A_y \delta{\vec{y}} + A_{8\times1} \delta x_9.
\label{reduced}
\end{equation}

The angular momentum magnitude integral, $H$, is linearized and written as,
\begin{equation}
\frac{\partial H}{\partial \vec{X}} \delta \vec{X} = 0,
\end{equation}
where the angular momentum vector is defined as,
\begin{equation}
\vec{H} = \mathcal{I}\vec{\omega}+\nu\vec{r}\times(\dot{\vec{r}}+\vec{\omega}\times\vec{r}).
\end{equation}

This can be expanded by again splitting the state:
\begin{equation}
\frac{\partial H}{\partial \vec{y}}\delta\vec{y} + \frac{\partial H}{\partial x_9}\delta x_9 = 0.
\end{equation}

Finally we can write:
\begin{equation}
\delta x_9 = \frac{\partial H}{\partial \vec{y}}\delta\vec{y}\left(\frac{-\partial H}{\partial x_9}\right)^{-1}.
\end{equation}
Substituting this in gives,
\begin{equation}
\delta{\dot{\vec{y}}} = A^*\delta\vec{y},
\end{equation}
with,
\begin{equation}
\label{eq: A*}
A^* = A_y - A_{8\times1}\frac{\partial H}{\partial \vec{y}}\left(\frac{\partial H}{\partial x_9}\right)^{-1}.
\end{equation}

At an equilibrium point it becomes possible to calculate the $A^*$ matrix and in turn find its spectral decomposition, with the zero eigenvalue corresponding to $\omega_3$ removed. The presence of real components in any of the eigenvalues of $A^*$ would correspond to unstable motion. Furthermore, the eigenvalues can be leveraged to find resonances between the system's fundamental frequencies. Using this approach, we can compute the fundamental frequencies of the secondary's motion as a function of its axial ratios $a/b$ and $b/c$. The results of this analytic approach will be described later in Section \ref{subsection: analyticResults}.

\subsection{Simplified 3D Dynamics Model}
\label{subsection: simplified3Dmomdel}
The ``simplified 3D model'' is an efficient approximation of the mutual spin-orbit dynamics that captures Dimorphos's libration behavior. First, the mutual orbit is integrated based on the equations of motion in which the mutual potential is expanded to second-order, accounting for the primary's $J_{2}$ moment and the secondary's ellipsoidal shape \citep{McMahon2013}. Although the full shape model of Didymos is not used, its $J_{2}$ moment alone is a reasonable approximation due to its fast rotation. The equations of motion for the orbit are described in detail in Appendix \ref{appendix: J2+ell}.  

The 3D spin and attitude of the secondary are then integrated via Euler's rigid-body equations for a triaxial ellipsoid \citep{Wisdom1984}, using the mutual orbit found in the previous integration (see Appendix \ref{appendix: 3D attitude}). It is important to note that for small variations from the planar solution (small obliquity of the spin axis) the rotation of Dimorphos matches the planar one very closely. Only when the precession of the spin axis is significantly excited does the model fail to produce the correct 3D spin-orbit coupled motion, due to the lack of conservation of the total angular momentum of the system. However, because it is unlikely that the DART impact will induce an \textit{immediate} large-amplitude precession in the secondary, this simplified approach lends itself to being an extremely efficient way of studying the secondary's attitude dynamics, over a wide range of possible shapes and other parameters. In any case, this approach is valid for small deviations from the planar case or short-term integrations and is sufficient for deducing the attitude stability properties under perturbations. For this purpose, it is necessary to derive also the linearized (variational) equations of the system, which are integrated simultaneously with the equation of motion, to derive the stability properties (see Appendix \ref{appendix: 3D attitude}).

\subsection{The GUBAS Full-Two-body-Problem code}
\label{subsection: GUBAS}
The General Use Binary Asteroid Simulator (GUBAS) is a novel F2BP code that uses the inertia integral method for evaluating the mutual potential between two arbitrary rigid bodies. The mathematical formulation for inertia integrals is described in \cite{hou2017} and implemented in a fast, open-source\footnote{The code is available at https://github.com/alex-b-davis/gubas} \textsc{c++} code with a Python-based user interface \citep{davis2020}. The code has several options for integration scheme, body-shape representations, and gravity expansion order. In the results presented here, we use the Lie group variational integrator and a fixed timestep of 40.0 seconds, which has been shown to give numerically converged results for this system \citep{agrusa2020}. The primary is represented by its radar-derived shape model \citep{naidu2020}, the secondary is a triaxial ellipsoid with adjustable axial ratios, and the mutual gravity is expanded to 4$^{\text{th}}$ order. The equations of motion are then integrated, with the mutual orbit and body spins fully coupled. See \cite{davis2020} for more details on this code. 

\subsection{Problem Set Up}
\label{subsection: probSetUp}
In this paper, we are exploring the binary orbital evolution solely under the influence of the mutual gravitational potential. Both bodies are considered to be fully rigid with the same bulk density. This study considers much shorter timescales than those associated with higher-order perturbations such as mutual tides, YORP, BYORP, and solar gravity, which are ignored here. In the simulations presented in this work, we use an integration time of one year. This timescale is long enough to allow for any strong attitude instabilities to set in, but short enough that higher-order perturbations to the mutual orbit can be safely ignored. 

We assume that the mutual orbit is initially planar, with the spin poles of both bodies aligned with the mutual orbit pole. Further, we assume the pre-impact orbit is nearly circular and that the secondary is in the 1:1 spin-orbit resonance with the libration amplitude damped to a minimum. We adopt the latest observed parameters from the DART Design Reference Asteroid (DRA), namely, the primary and secondary sizes, the binary semimajor axis, and the binary orbit period. These parameters are listed and referenced in Table \ref{table: params}. In order to achieve the assumptions listed above, and to match the observed DRA parameters (namely the measured orbit period, which has been measured to high precision), we use an optimization scheme to determine our initial conditions for the GUBAS simulations\footnote{This procedure is only required for GUBAS, as it uses the polyhedral shape model for the primary and evaluates the mutual gravity to 4$^{\text{th}}$ order. However, this procedure can be done analytically for the simplified 3D model (see Appendix \ref{appendix: J2+ell}).}. A naive approach using Kepler's 3$^{\text{rd}}$ law to derive the mass of the system is invalid due to the non-spherical shapes and close proximity of the two components. Therefore, our initial-conditions-optimization scheme adjusts the total mass of the system (assuming the primary and secondary have the same bulk density), until it finds a mass where the simulated orbit period matches the observed period. This process is able to generate initial conditions that match the observed orbit period to high precision that also have a small libration amplitude and nearly circular orbit. This means that when the shape of the secondary is changed (although its total volume is conserved), the system mass and bulk density change slightly. The mass and density adjustments are small ($<1\%$) and allow us to match the observed orbit period to the highest precision possible because the orbit period is so well constrained. The details of this optimization routine can be found in Appendix \ref{appendix:optimization scheme}. 

For each choice of the secondary's axis ratios $a/b$ and $b/c$, the optimization scheme is used to derive the pre-impact relaxed state of the system. Then, the orbital speed of the secondary is altered according to our choice for $\beta$. Based on the most recent DART flight plans at the time of this writing, the choice for spacecraft mass and relative speed were $535 \text{ kg}$ and $6.6\text{ km s}^{-1}$, respectively. These values are subject to change by small amounts, but are not expected to change drastically. With the head-on, planar impact considered in this work, Eq.\ \eqref{eq: deltaV_vector} can be simplified to give the perturbation to the secondary's orbital speed:

\begin{equation}
	\label{eq: deltaV}
    \Delta v = -\beta\frac{ M_{\text{DART}}v_{\text{DART}}}{M_{\text{B}}},
\end{equation}
where $ M_{\text{DART}}$ and $v_{\text{DART}}$ are the respective mass and speed of the DART spacecraft, and $M_{\text{B}}$ is the mass of Dimorphos. The change in speed is negative because the nominal impact trajectory impacts the leading face of Dimorphos (head-on impact), causing it to slow down, fall onto a tighter orbit, and reduce the orbit period. The real DART trajectory will result in an impact with Dimorphos at an angle relative to the mutual orbit plane that varies with launch date within a range of roughly 5--30 degrees. However, in this work, we assume an idealized head-on impact with no out-of-plane component. We also assume that the impact is aligned with the center of mass, such that there is no instantaneous torque imparted to Dimorphos. We leave the more realistic treatment of the impact geometry to future work. We note that a non-planar and off-center impact will likely excite the secondary's spin state significantly more than in an idealized, head-on impact. Therefore, the results presented in this work may be \textit{underestimating} the perturbation to Dimorphos's spin state.
\begin{table}
\centering
    \begin{minipage}{\textwidth}
    \begin{center}
        \begin{tabular}{ l l l }
            \bf\normalsize Parameter & \bf\normalsize Value  & \bf\normalsize Reference(s) \\ 
            \toprule
            Diameter of Primary $D_{\text{P}}$ & $780\pm30\text{ m}$ & \cite{naidu2020} (equivalent spherical diameter)\\
            \midrule
            Diameter of Secondary $D_{\text{S}}$ & $164\pm18\text{ m}$  &\cite{naidu2020, scheirich2009} \\
            \midrule
            Semi-major Axis $a_{\text{orb}}$& $1.19\pm{0.03}\text{ km}$  & \cite{naidu2020} \\
            \midrule
            Binary Orbit Period $P_{\text{orb}}$ & $11.9217\pm0.0002\text{ h}$ & Scheirich, P., personal communication (2020)\footnote{The best available orbital solution at the time of this work.} \\
            \midrule
            Primary Spin Period $P_{\text{P}}$ & $2.2600 \pm 0.0001 \text{ h}$ & \cite{pravec2006} \\
            \midrule
        \end{tabular}
        \end{center}
    \caption{\label{table: params} Physical parameters of the Didymos binary based on lightcurve and radar observations. We assume that the binary eccentricity and inclination are both zero. The initial conditions for the simulations presented here match all of these parameters, and all other initial conditions (i.e., masses and velocities) are derived from these parameters. }
    \end{minipage}
\end{table}

\section{Results}
\label{section: results}

\subsection{Analytic Model Results}
\label{subsection: analyticResults}

We performed a grid search over the solution space of axis ratios, ranging from $1<a/b<1.5$ and $1<b/c<1.5$. Due to the lack of a well-constrained shape for Dimorphos, the parameter space was instead selected because of an observed upper-limit of binary asteroid satellites with elongations $a/b > 1.5$ in the near-Earth, Mars-crossing, and small main belt populations \citep{pravec2016}. For each value of $a/b$ and $b/c$, the inertia tensor is computed for a uniform triaxial ellipsoid and normalized. The dynamics matrix, $A^*$ (Equation \eqref{eq: A*}), is then evaluated for each value of $a/b$ and $b/c$ at their respective equilibrium points. We find that all eigenvalues over this solution space are purely imaginary, which indicates stable motion about the equilibrium point.

\begin{figure}[!h]	
   \centering
   \includegraphics[width = 0.75\textwidth]{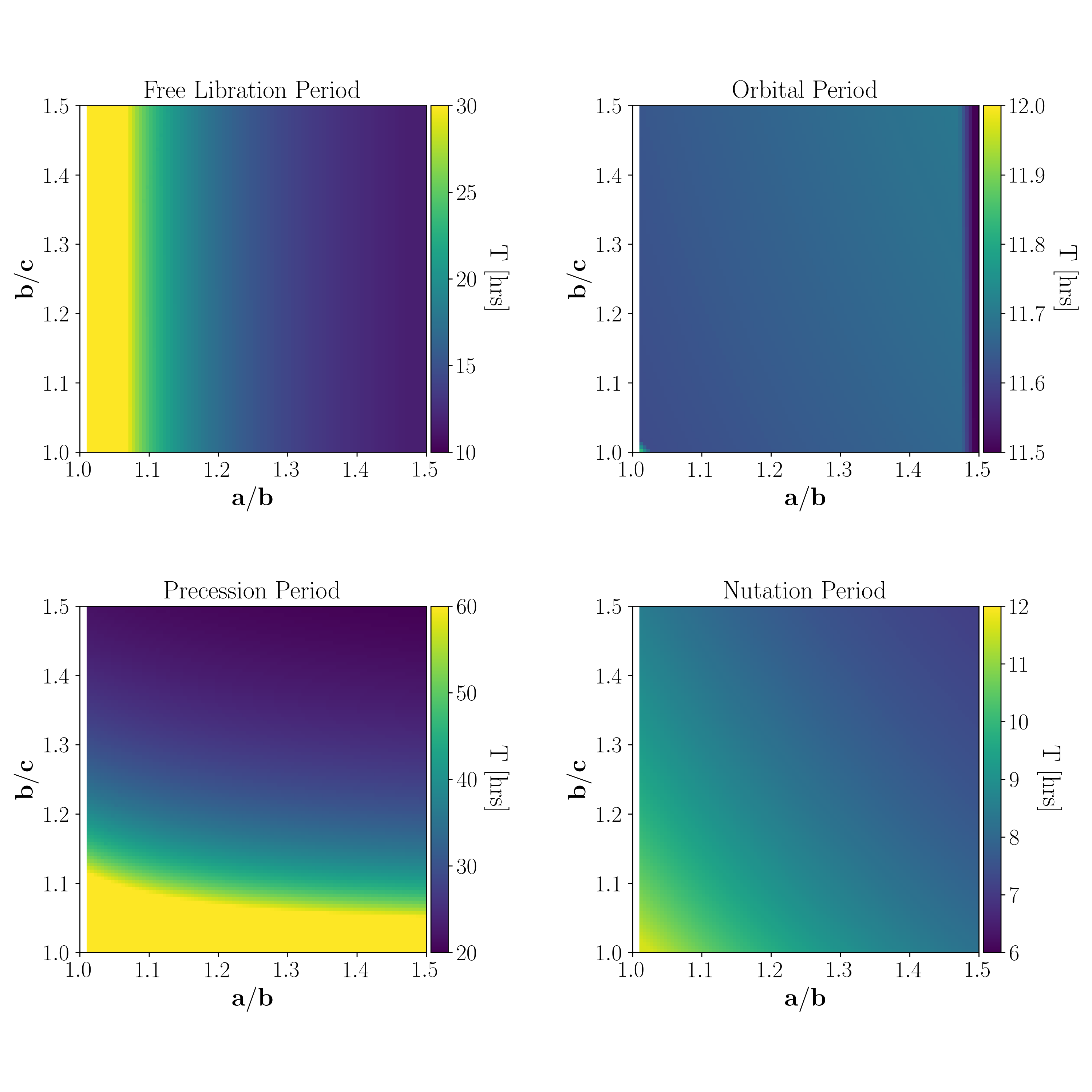} 
   \caption{The four fundamental periods of motion about the equilibrium point, shown in hours. The free libration period is very similar to the frequency given in Eq.\ \eqref{eq: librationFrequency}. The two nonplanar periods correspond to the secondary's spin precession and nutation periods.}
   \label{fig: fundamentalFrequencies}
\end{figure}

Because the spectral decomposition produces eight conjugate frequencies, there are only four unique values leading to four fundamental frequencies. The four fundamental frequencies from this analysis are represented by their period in Fig.\ \ref{fig: fundamentalFrequencies}. These four frequencies correspond to the in-plane free libration (i.e., Equation \eqref{eq: librationFrequency}), the orbital frequency (mean motion), and two out-of-plane frequencies, related to the precession and nutation of the secondary. 

\begin{figure}[!h]	
   \centering
   \includegraphics[width = 0.6\textwidth]{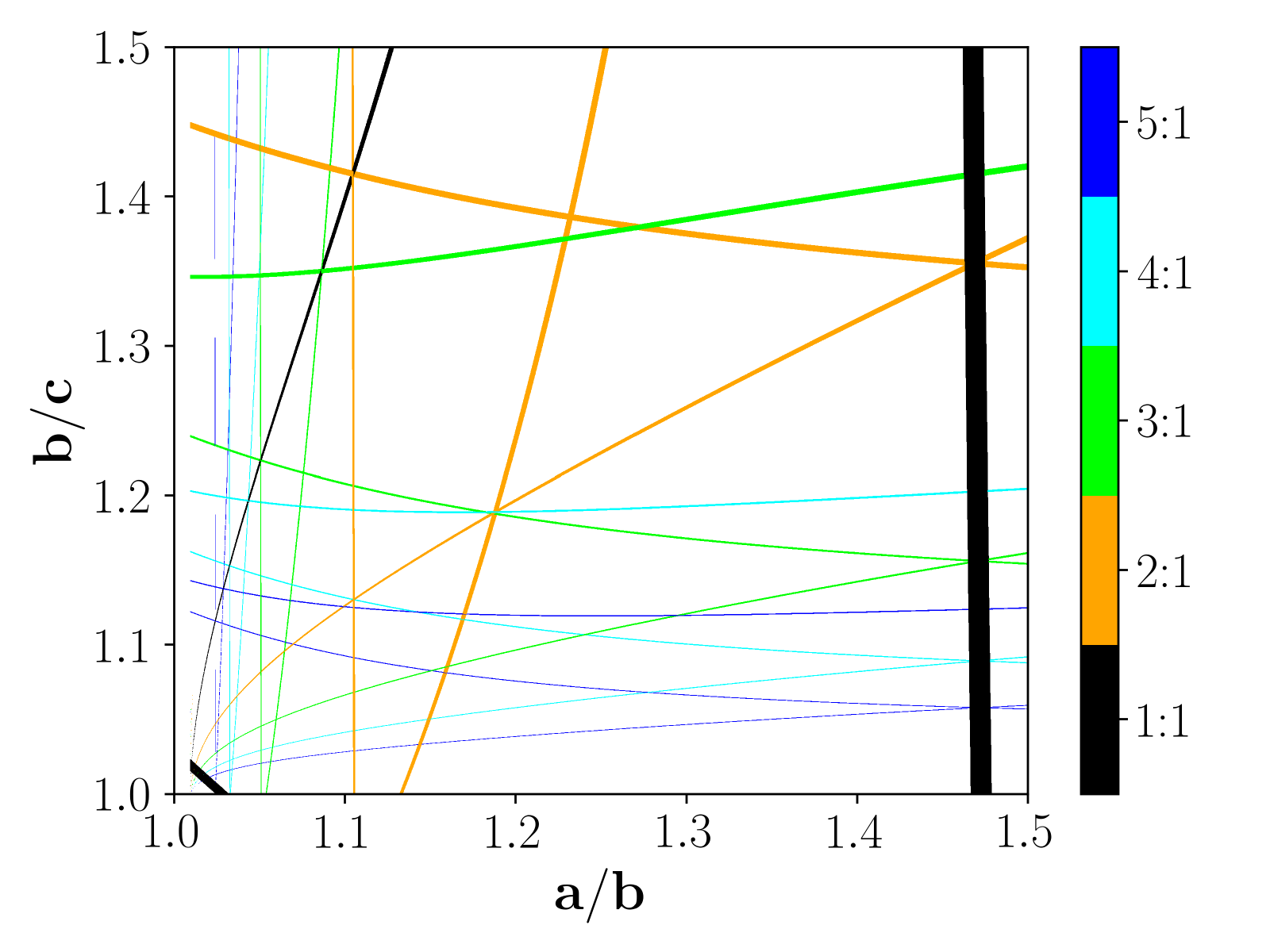} 
   \caption{The resonances of the four fundamental frequencies up to 5:1. The line weight is a reflection of the width of each resonance region, although each resonance is ideally a line. Thus, thicker lines allow for a wider range of secondary shapes to excite that resonance. Note that these lines are all continuous, and any lines that look dotted or dashed are due to the resonance width being smaller than the resolution of the grid search.}
   \label{fig: analyticResonances}
\end{figure}

Although the eigenvalue analysis naively indicates stable motion about the equilibrium (due to imaginary eigenvalues), we find a multitude of resonances among the fundamental frequencies upon closer examination . The resonance locations can be found by simply searching for locations in the solution space where one fundamental frequency becomes commensurate with another. These resonances indicate areas in the solution space in which this linear model is no longer accurate, and nonlinear effects become important. Figure \ref{fig: analyticResonances} shows each resonance between the various fundamental frequencies up to 5:1. It will turn out that some of these resonances will drive unstable motion in the full nonlinear problem, with the single 1:1 and three of the 2:1 resonances being the most dominant.

\newpage
\subsection{Simplified 3D Dynamics Results}
To be clear, in the previous section, we found the fundamental frequencies and resonance locations for Dimorphos at equilibrium (i.e., a uniform, circular orbit). Then, with the simplified 3D model (and later with GUBAS), we add a perturbation to Dimorphos's relative velocity according to a given choice for $\beta$ to study the post-impact attitude stability. The results of Section \ref{subsection: analyticResults} do not depend on $\beta$ and only apply to the case the pre-impact circular orbit. However, Section \ref{subsection: analyticResults} provides insight for understanding the mechanisms that drive attitude instability when Dimorphos's orbit is perturbed.

\begin{figure}[t!]
    \centering
    \begin{subfigure}[t]{0.5\textwidth}
        \centering
        \includegraphics[width=\textwidth]{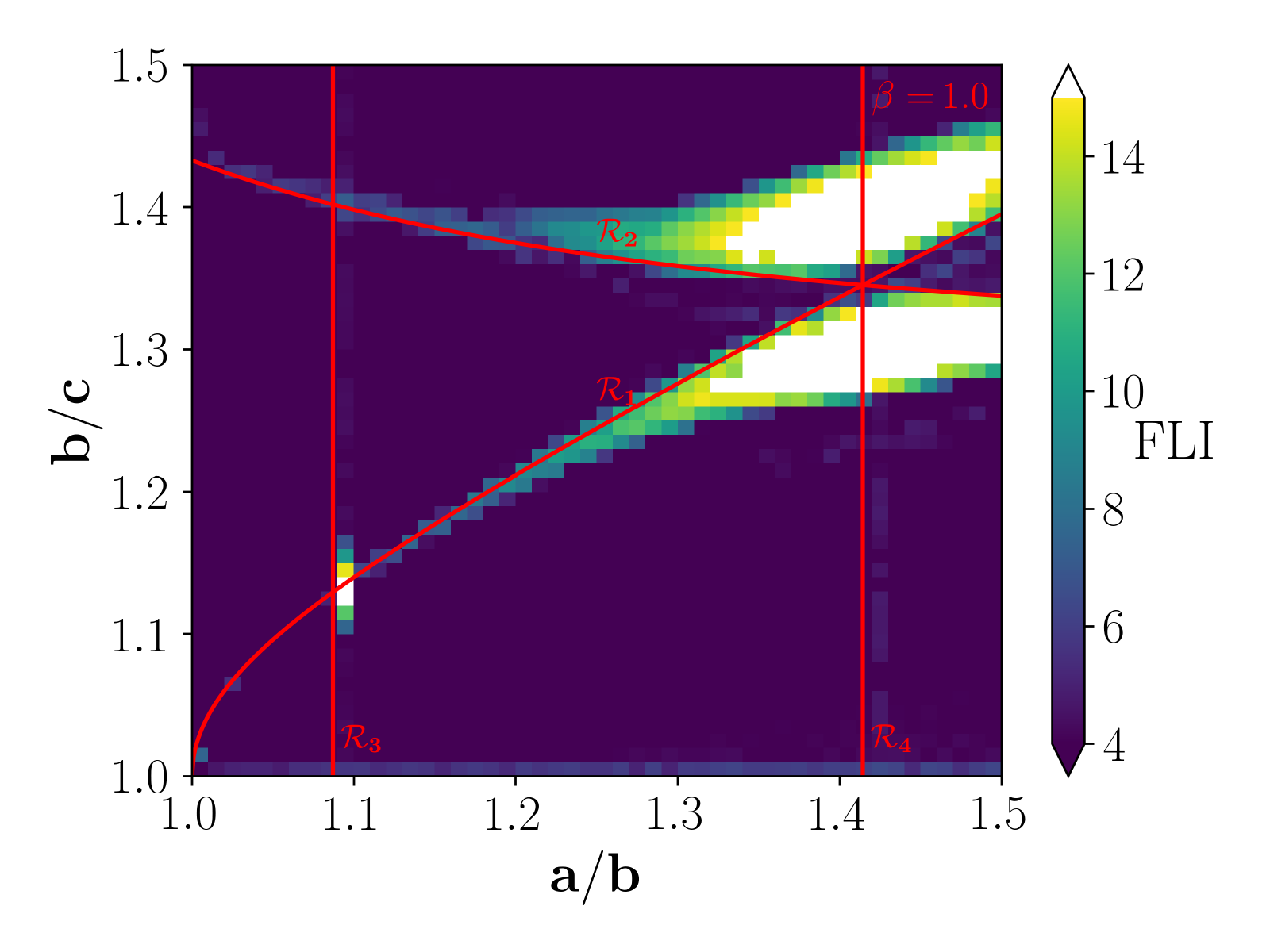}
        \caption{$\beta = 1$\label{fig: FLI_beta1}}
    \end{subfigure}%
    ~ 
    \begin{subfigure}[t]{0.5\textwidth}
        \centering
        \includegraphics[width=\textwidth]{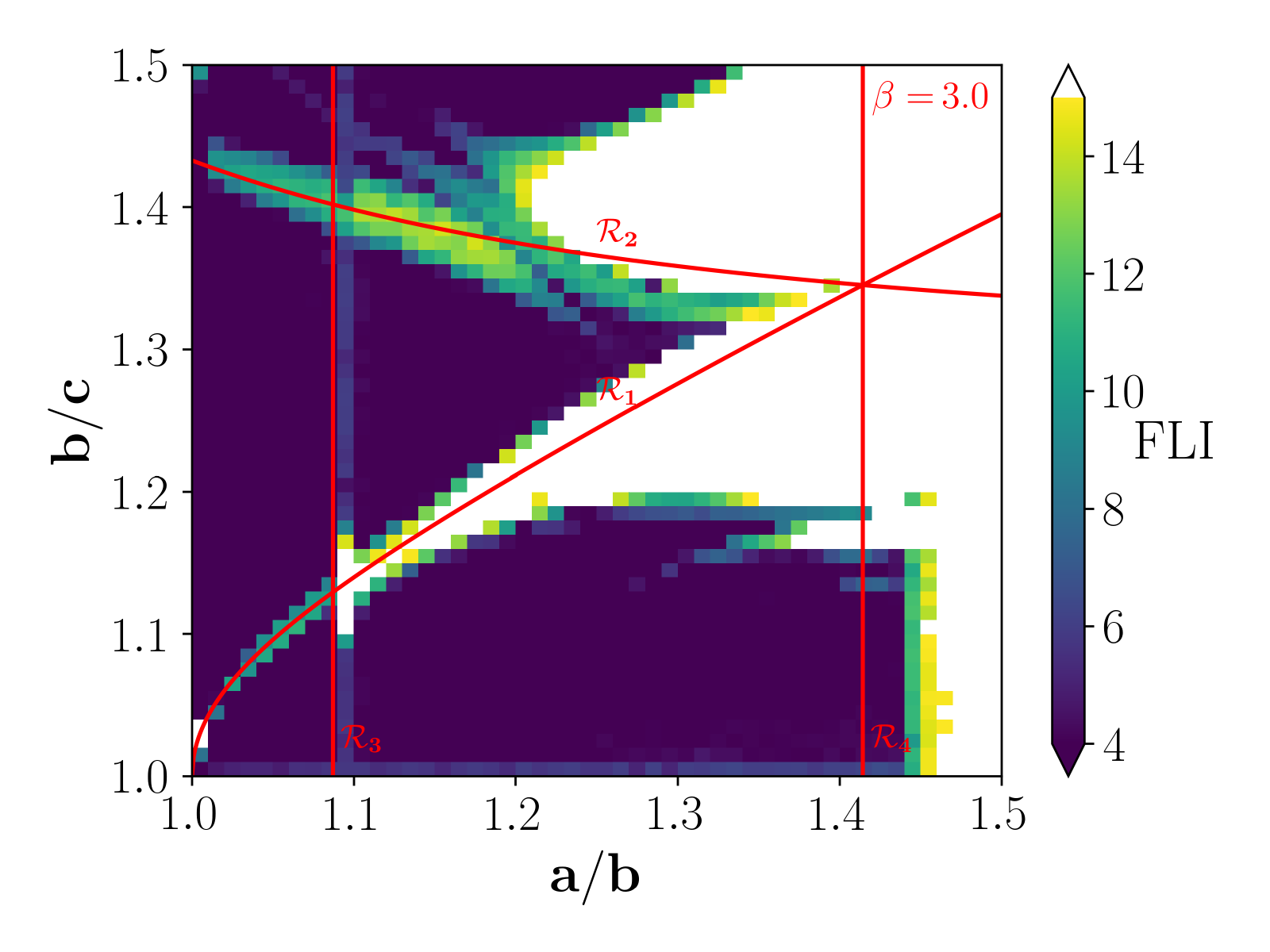}
        \caption{$\beta = 3$\label{fig: FLI_beta3}}
    \end{subfigure}
  \caption{\label{fig: FLI} A fast Lyapunov indicator map of the parameter space, with larger values indicating unstable, chaotic motion in the secondary's spin state. The uncoupled resonance locations from Table \ref{table: resonances} are overlaid to show the dominant drivers of instability.}
\end{figure}

Using the simplified 3D model, we first computed the fast Lyapunov indicator (FLI) for each combination of $a/b$ and $b/c$ using the simplified 3D model for $\beta=1$ and $\beta=3$. The FLI is a useful and widely used tool for detecting weak chaos in dynamical systems \citep{Froeschle1997}, and is a measure of the exponential divergence in phase space of two solutions with infinitesimally separated initial conditions. The resulting FLI map is shown in Fig.\ \ref{fig: FLI}. It should be noted that a run for $\beta = 0$ (i.e., the pre-impact, relaxed state) yields a fully regular phase space. This has further been confirmed via a Floquet analysis of the relaxed states, which showed that all computed eigenvalues indicated stable motions. Moreover, this finding is in agreement with the analytical approach of Section \ref{subsection: analyticResults}. 

Based on the shape of the instability region identified in the FLI analysis, the chaotic motion seems to be primarily driven by four key resonances, which are given on Table \ref{table: resonances}. It should first be noted that the frequencies and resonances given in Table \ref{table: resonances} are the uncoupled frequencies described in Appendix \ref{appendix: natural frequencies}. In reality, the true frequencies and resonance locations are slightly different (and more complicated) and the uncoupled frequencies are only meant to be a qualitative indicator here. In this case, the uncoupled resonances do an adequate job in predicting the locations of resonances.

The three prominent frequencies seem to be the mean motion, $n^{uc}$, the free libration frequency, $\omega^{uc}_{\text{lib}}$, and the secondary's spin precession frequency, $\omega^{uc}_{\text{prc}}$. These frequencies have the superscript $uc$ to indicate that they are \textit{uncoupled} and merely an approximation to the real frequency. To first order, the free libration frequency is approximated by Eq. \eqref{eq: librationFrequency}. For a uniform triaxial ellipsoid, the principal moments of inertia can be rewritten in terms of the corresponding semi-axis lengths $a$, $b$, and $c$. The libration frequency, $\omega^{uc}_{\text{lib}}$ is in a 1:1 resonance with the mean motion when $a/b = \sqrt{2}$ and a 2:1 resonance when $a/b = \sqrt{13/11}$. These resonances appear as two faint vertical lines on Fig.\ \ref{fig: FLI_beta1} and are referred to as $\mathcal{R}_{4}$ and $\mathcal{R}_{3}$, respectively, in Table \ref{table: resonances}.

The secondary's spin precession frequency, $\omega^{uc}_{\text{prc}}$, is more complicated and is given in Appendix \ref{appendix: natural frequencies}. For certain combinations of $a/b$ and $b/c$, $\omega^{uc}_{\text{prc}}$ can enter a 2:1 resonance with $\omega^{uc}_{\text{lib}}$ or a 2:1 resonance with $n$. These two resonances make up the two wing-like structures in Fig.\ \ref{fig: FLI_beta1} and are called $\mathcal{R}_{1}$ and $\mathcal{R}_{2}$, respectively, in Table \ref{table: resonances}. These four resonances among the three frequencies are certainly not the only ones playing a role in the structure of the instability region, but seem to be the dominant contributors. 

\begin{figure}[t!]
    \centering
    \begin{subfigure}[t]{0.5\textwidth}
        \centering
        \includegraphics[width=\textwidth]{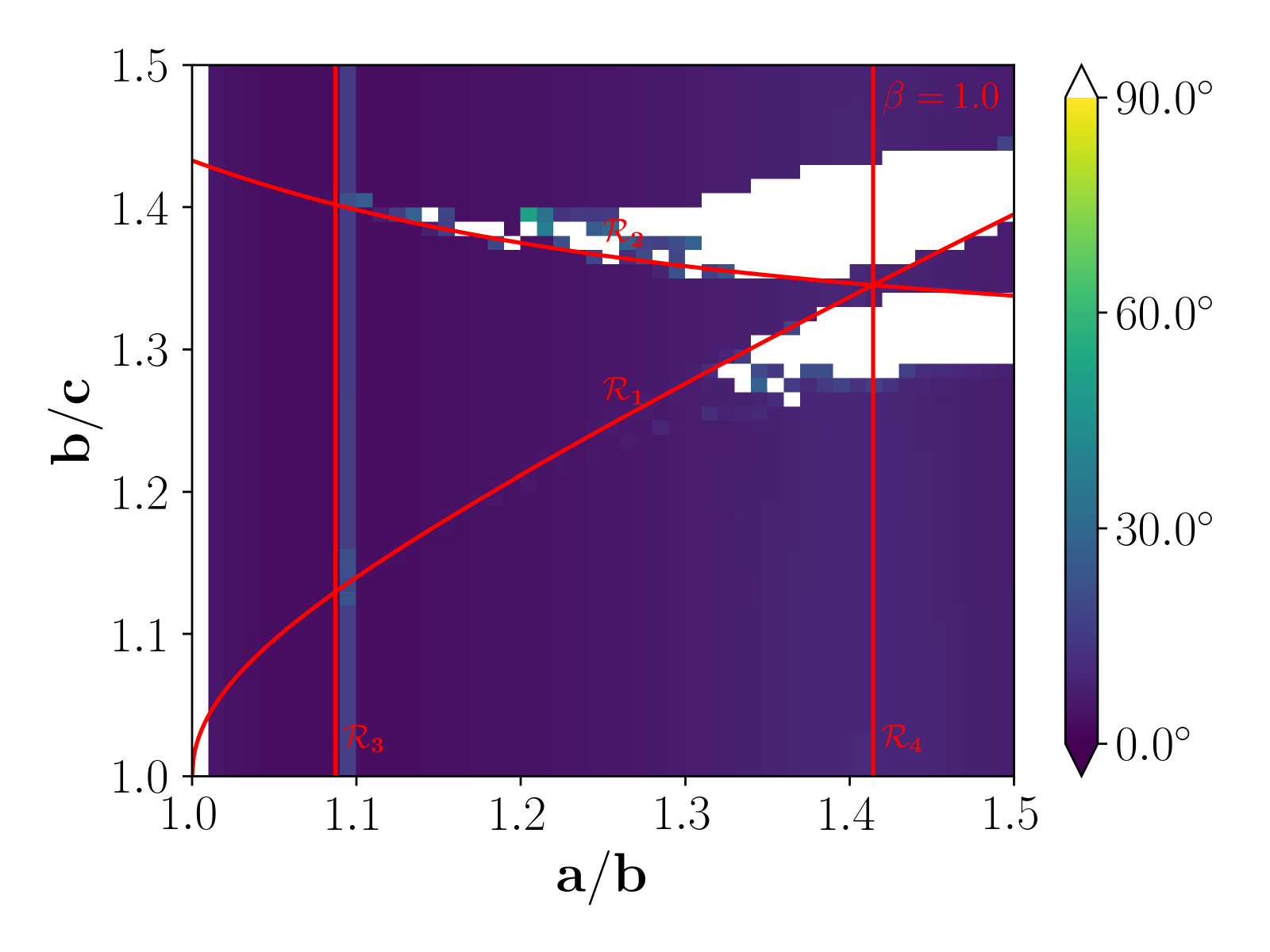}
        \caption{$\beta = 1$\label{fig: planarLibration_beta1}}
    \end{subfigure}%
    ~ 
    \begin{subfigure}[t]{0.5\textwidth}
        \centering
        \includegraphics[width=\textwidth]{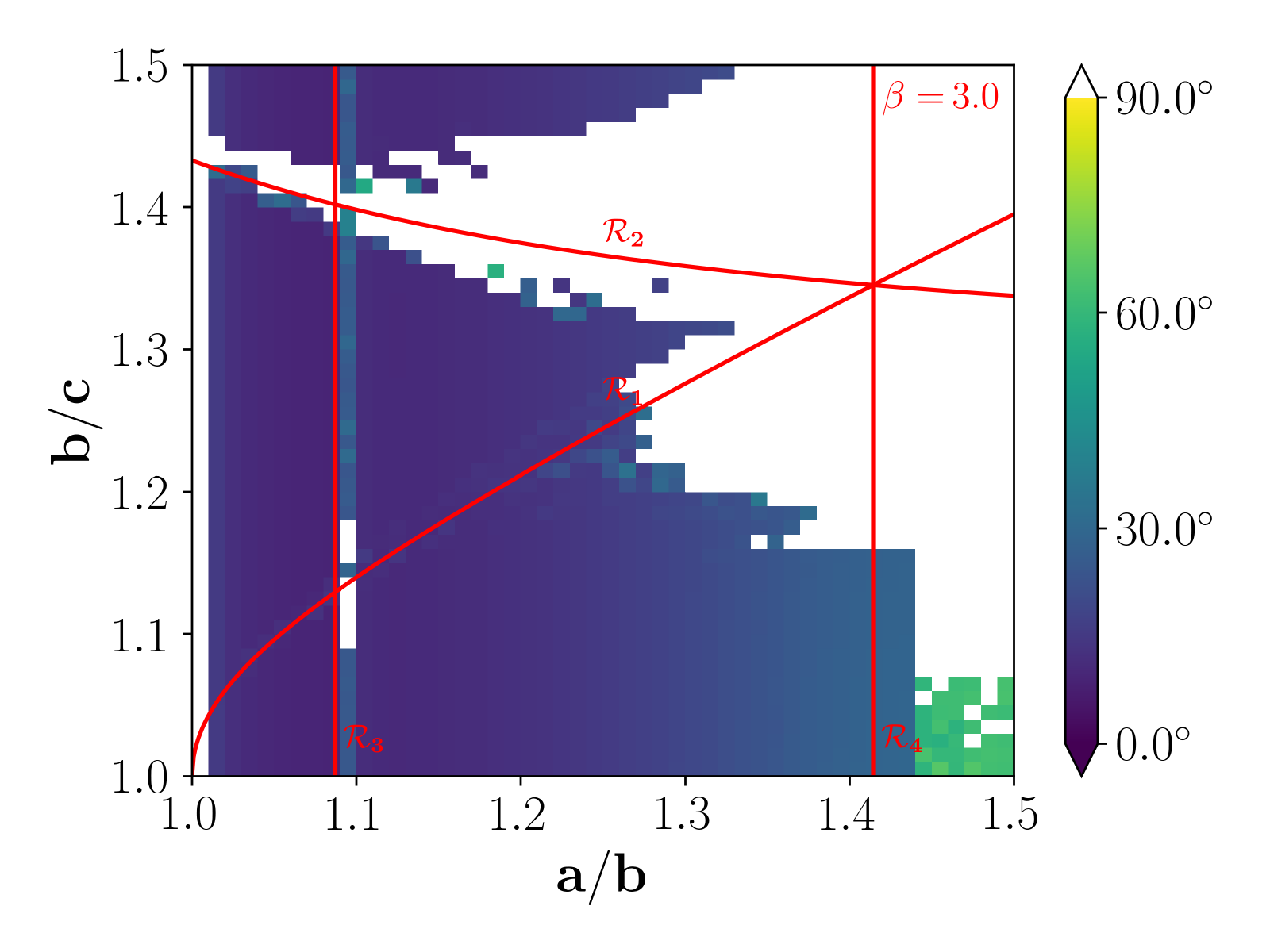}
        \caption{$\beta = 3$\label{fig: planarLibration_beta3}}
    \end{subfigure}
    \caption{\label{fig: planarModel} The maximum libration amplitude for $\beta = 1$ and $\beta = 3$ as a function of the secondary's axial ratios $a/b$ and $b/c$, according to the simplified 3D model. When the libration angle exceeds $90^{\circ}$, the satellite has broken from synchronous rotation. The \textit{uncoupled} resonance locations from Table \ref{table: resonances} are overlaid to show the dominant drivers of instability.}
\end{figure}

The maximum libration angle achieved for each secondary shape after a one-year simulation is shown in Fig.\ \ref{fig: planarModel}. This angle is not necessarily entirely within the orbit plane and can have nonplanar components that we investigate later. When the libration angle exceeds $90^{\circ}$, we consider Dimorphos to have broken from synchronous rotation, which is shown in white on the plot. Even for $\beta=1$, we see there are a significant number of cases in which the libration angle has exceeded $90^{\circ}$, and when $\beta=3$, nearly half of the parameter space has exceeded $90^{\circ}$. We find that the cases where the libration amplitude exceeds $90^{\circ}$ are largely correlated to chaotic motion identified by the FLI. The four resonances identified in the FLI analysis are overlaid onto the libration plots to show the qualitative agreement between the analytic, uncoupled resonance locations and the instability regions. However, when the secondary's spin state becomes significantly excited (i.e., large nonplanar oscillations), the simplified 3D model is no longer valid, requiring the use of a fully coupled solver.

\begin{table}
    \begin{tabular}{l | c}
      Name & Resonance \\ \hline
      $\mathcal{R}_{1}$ & $\omega^{uc}_{\text{lib}} = 2\omega^{uc}_{\text{prc}}$ \\ 
      $\mathcal{R}_{2}$ & $n^{uc} = 2\omega^{uc}_{\text{prc}}$ \\ 
      $\mathcal{R}_{3}$ & $n^{uc} = 2\omega^{uc}_{\text{lib}}$ \\ 
      $\mathcal{R}_{4}$ & $\omega^{uc}_{\text{lib}} = n^{uc}$ \\ 
    \end{tabular}
  \caption{\label{table: resonances} The four main resonances driving Dimorphos's attitude instability. $\omega^{uc}_{\text{lib}}$ is the free libration frequency of the secondary, $n^{uc}$ is the mean motion, and $\omega^{uc}_{\text{prc}}$ is the spin precession frequency of the secondary. We use the superscript $uc$ to indicate that these frequencies are uncoupled and are only approximations of their real value in the fully coupled problem.}
\end{table}

\subsection{GUBAS Results}
\begin{figure}[t!]
    \centering
    \begin{subfigure}[t]{0.5\textwidth}
        \centering
        \includegraphics[width=\textwidth]{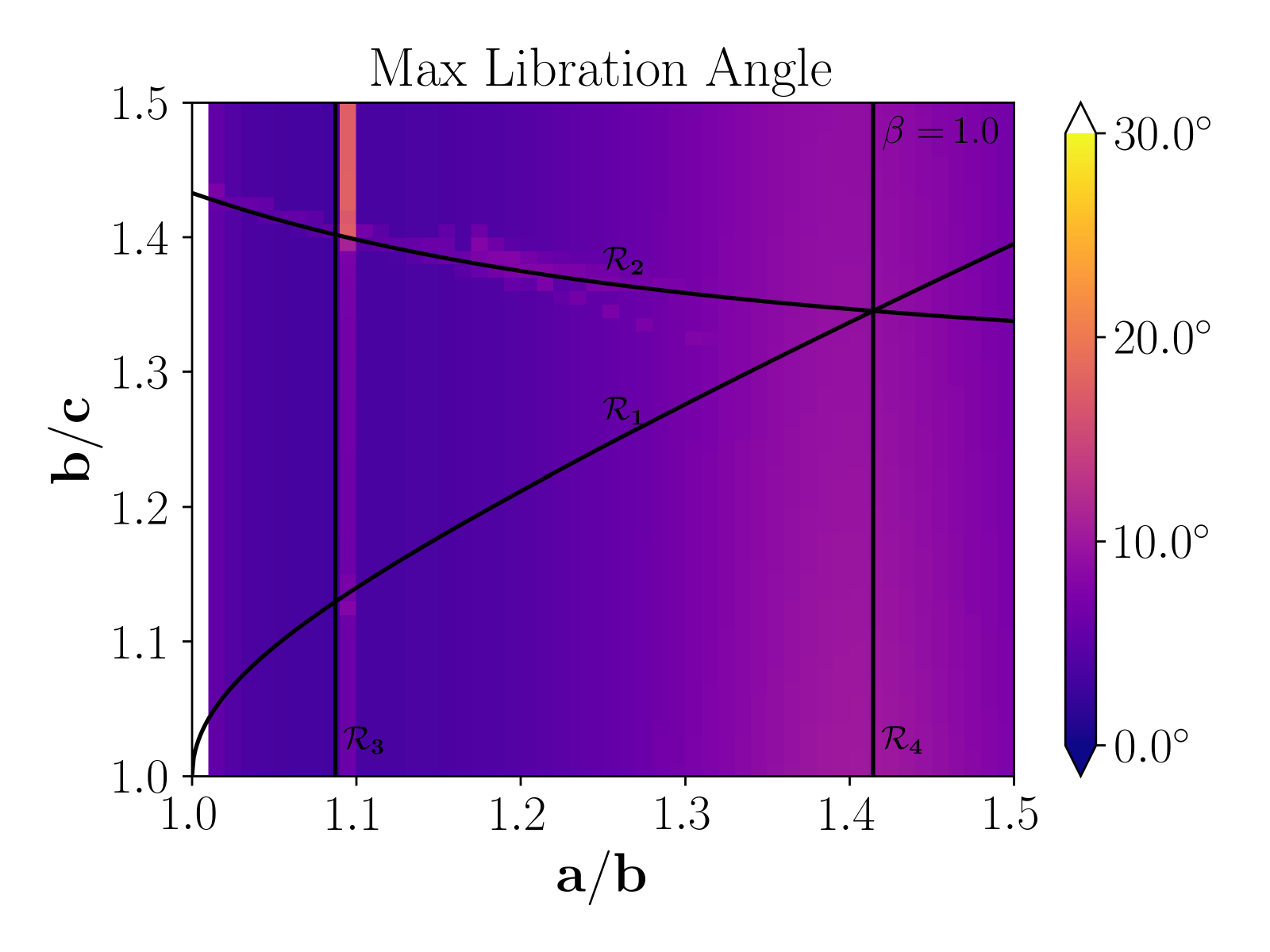}
        \caption{$\beta = 1$\label{fig: gubasLibration_beta1}}
    \end{subfigure}%
    ~ 
    \begin{subfigure}[t]{0.5\textwidth}
        \centering
        \includegraphics[width=\textwidth]{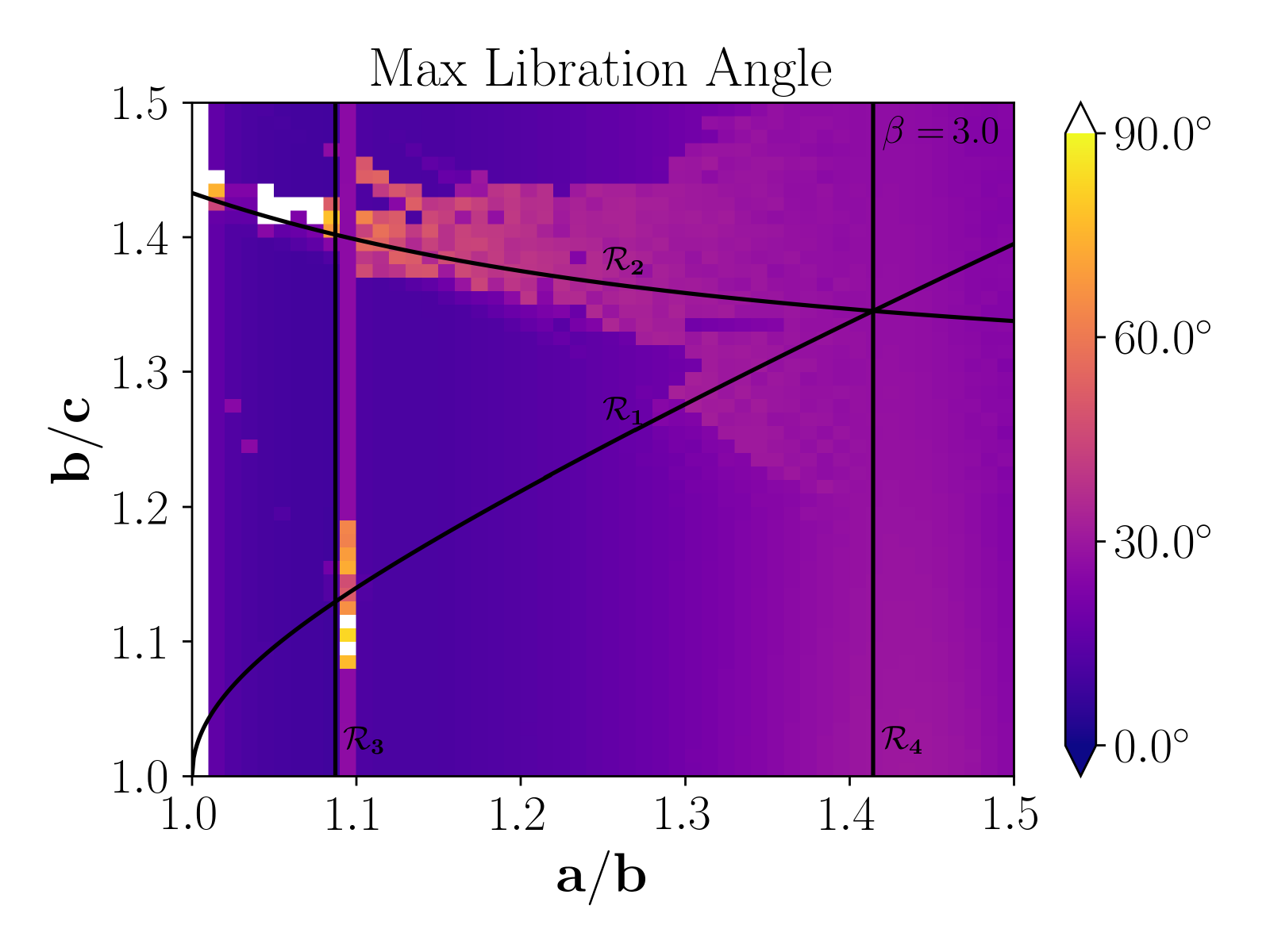}
        \caption{$\beta = 3$\label{fig: gubasLibration_beta3}}
    \end{subfigure}
  \caption{\label{fig: gubasLibration} The maximum libration amplitude in GUBAS simulations with the theoretical resonances from Table \ref{table: resonances} overlaid. We recover the same general dynamical structure seen in the simplified 3D model. However, the libration amplitudes are far smaller than those in the simplified model (Fig.\ \ref{fig: planarModel}), and only a few cases ever exceed the $90^{\circ}$ threshold for leaving synchronous rotation. Note the different colorbar scales for \ref{fig: gubasLibration_beta1} and \ref{fig: gubasLibration_beta3}.}
\end{figure}

Equivalent simulations were run in GUBAS over the full parameter space of $a/b$ and $b/c$ for $\beta$ values of 1, 3, and 5. The core differences between the two simulation codes is the use of the radar-derived polyhedral shape model for the primary, expansion of the mutual gravitational potential to higher order  (4$^{\text{th}}$), and full coupling between the body spins and mutual orbit. The maximum libration amplitude achieved over the course of the one-year simulations is shown in Fig.\ \ref{fig: gubasLibration} for $\beta=1$ and $\beta=3$. Again, the libration angle is considered to be the angle between Dimorphos's long axis and the LOC, which is not necessarily planar. When $\beta=1$, no cases exceed $90^{\circ}$ and only a select few do at $\beta=3$. The libration amplitudes are quite large, but much smaller than those in the simplified 3D model runs. This is a direct result of Dimorphos traveling on a fully coupled orbit rather than the predetermined one in the simplified model. Put simply, the secondary transfers excess spin energy to the mutual orbit, keeping the libration amplitude from exceeding $90^{\circ}$. The libration amplitudes are still relatively large however, and given the chaotic nature of Dimorphos's tumbling state, it is possible that these cases could break from synchronous rotation after longer integration times. 

\subsubsection{Dimorphos's attitude instability}
\begin{figure}[t!]
    \centering
    \begin{subfigure}[t]{0.5\textwidth}
        \centering
        \includegraphics[width=\textwidth]{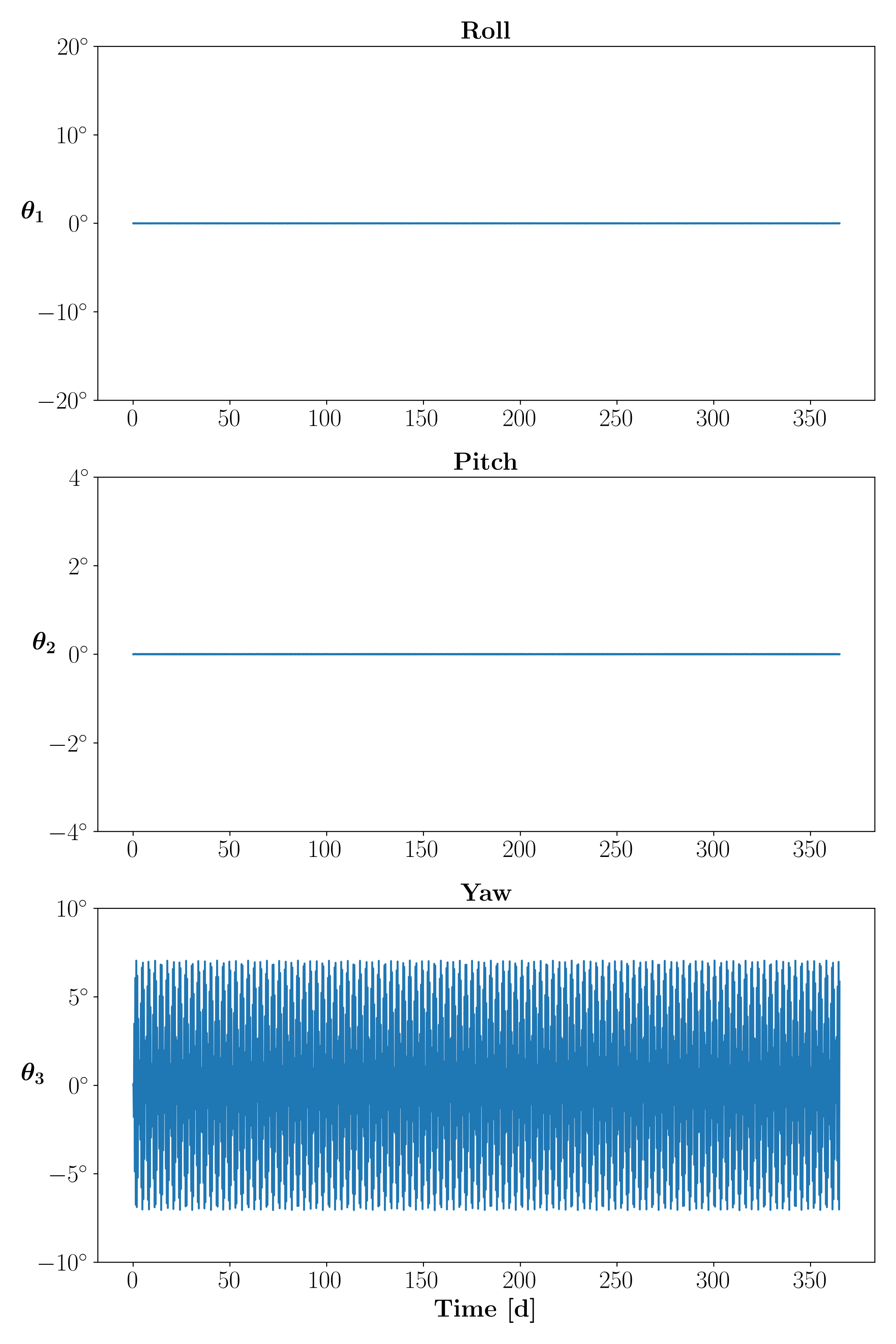}
        \caption{$\beta=1$, $a/b=1.3$, $b/c=1.2$\label{fig: eulerAngles_ab1.3bc1.2}}
    \end{subfigure}%
    ~ 
    \begin{subfigure}[t]{0.5\textwidth}
        \centering
        \includegraphics[width=\textwidth]{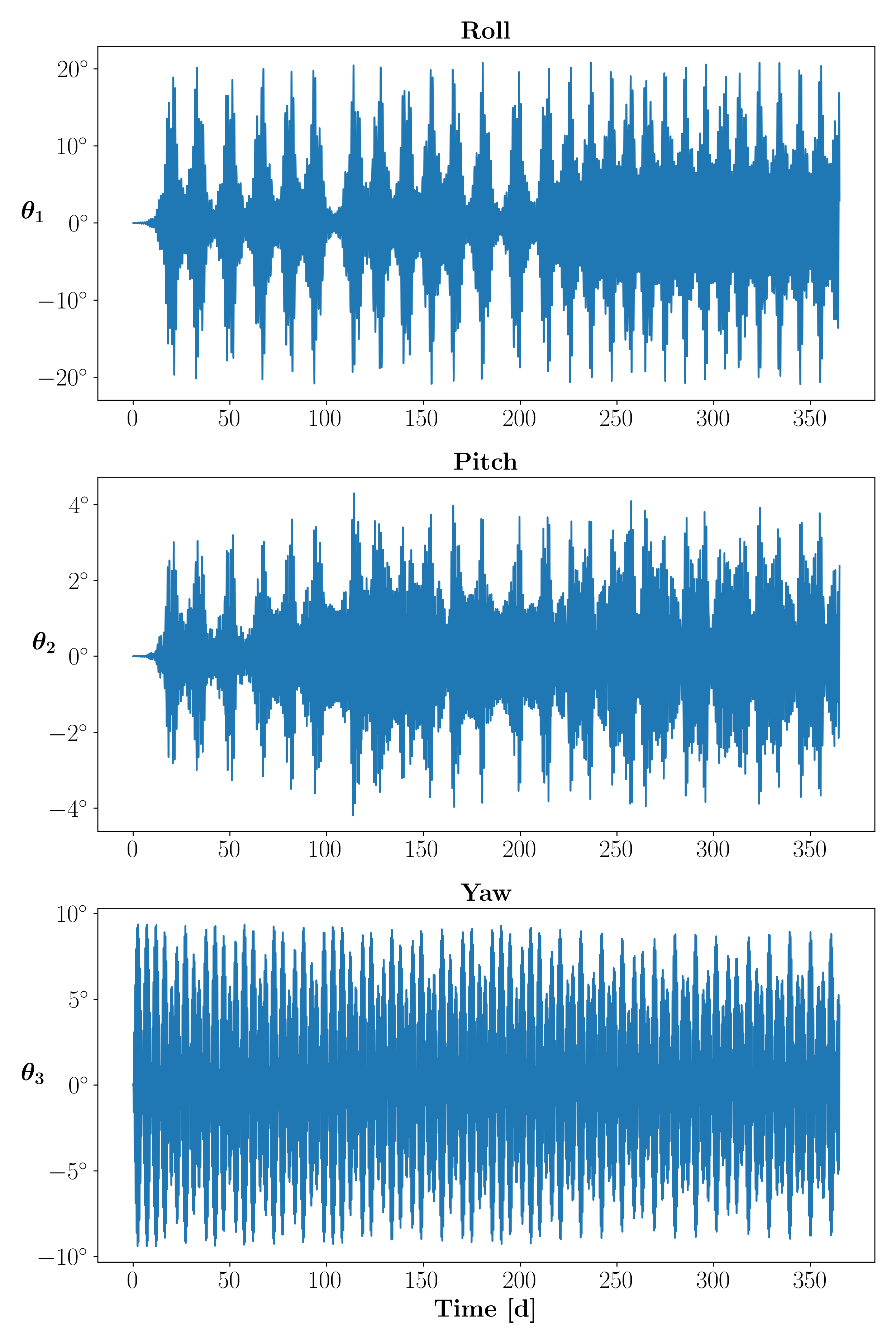}
        \caption{$\beta=1$, $a/b=1.4$, $b/c=1.3$\label{fig: eulerAngles_ab1.4bc1.3}}
    \end{subfigure}
    \caption{Time-series plots of the secondary's three Euler angles for two different secondary shapes: $a/b=1.3, b/c=1.2$ (a typical stable case) and $a/b=1.4, b/c=1.3$ (a typical unstable case). Note the difference in angle-axis scales for each Euler angle. The time axis shows the time since the DART impact (impact occurs at time $=0$).}
    \label{fig: eulerAngles_beta1}
\end{figure}

Instead of looking at just the libration angle defined as the angle between Dimorphos's long axis and the LOC, we can examine the 1-2-3 Euler angle set that make up the secondary's attitude as described in Section \ref{subsection: background}. Figure \ref{fig: eulerAngles_beta1} shows a time series of the three Euler angles for the simulation where $\beta=1$ for two different shapes of the secondary. Fig.\ \ref{fig: eulerAngles_ab1.3bc1.2} shows the Euler angles for a secondary in which $a/b=1.3$ and $b/c=1.2$, which represents a typical simulation in which the secondary's attitude remains stable. The roll and pitch angles remain small, hovering around zero, while the yaw angle steadily librates around an equilibrium. Changing the shape only slightly to $a/b=1.4$ and $b/c=1.3$, Fig.\ \ref{fig: eulerAngles_ab1.4bc1.3} shows a typical example where the secondary's attitude becomes unstable. Roughly ${\sim}10$ days after the DART impact, the secondary's roll and pitch angles become excited, reaching angles of ${\sim}20^{\circ}$ and ${\sim}4^{\circ}$ respectively. After the initial excitement of the roll and pitch angles, there is a continuous energy exchange between the planar and nonplanar degrees of freedom, indicated by the simultaneous changes in the roll and pitch angles.

\begin{figure}[t!]
    \centering
    \begin{subfigure}[t]{0.32\textwidth}
        \centering
        \includegraphics[width=\textwidth]{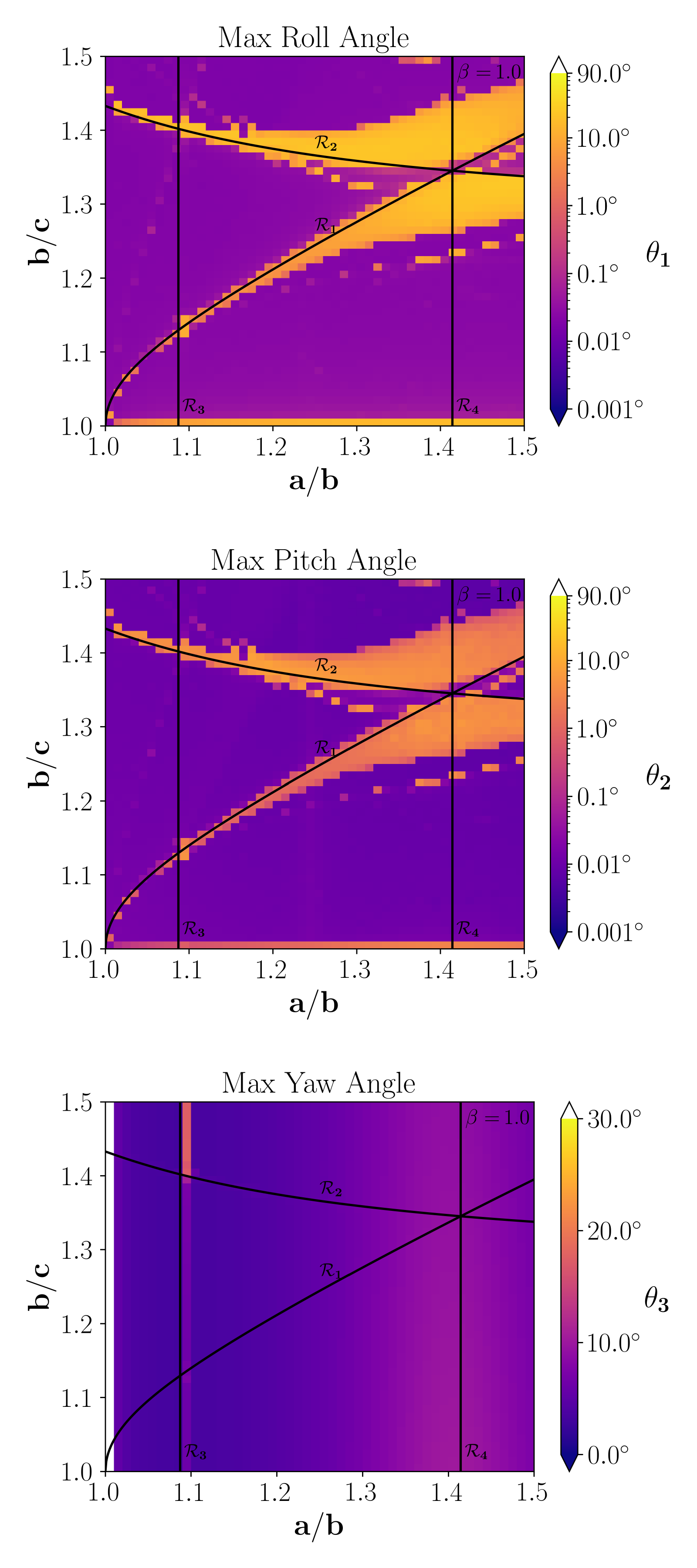} 
        \caption{Max Euler angles, $\beta=1$.}
        \label{fig: gubasEuler_beta1}
    \end{subfigure}%
    ~ 
    \begin{subfigure}[t]{0.32\textwidth}
        \centering
        \includegraphics[width=\textwidth]{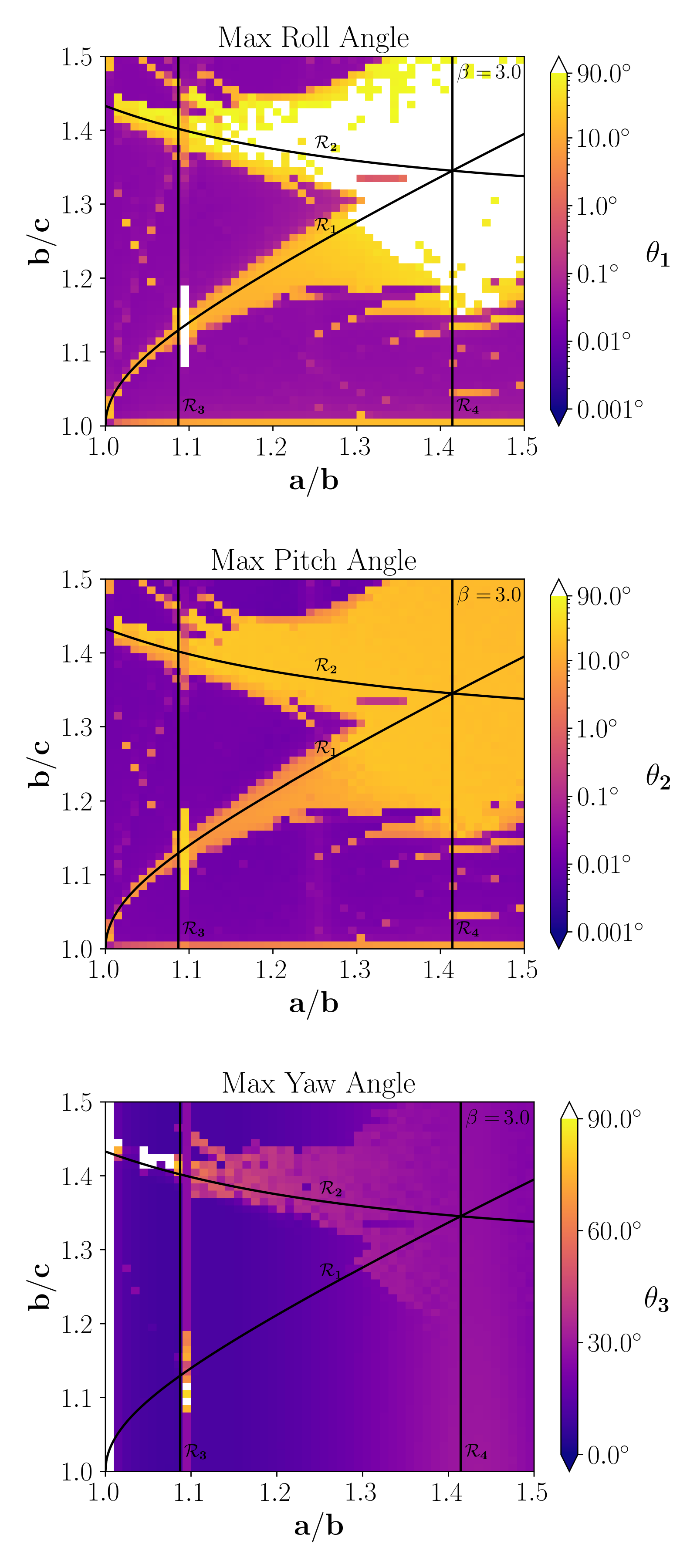} 
        \caption{Max Euler angles, $\beta=3$.}
        \label{fig: gubasEuler_beta3}
    \end{subfigure}
    ~ 
    \begin{subfigure}[t]{0.32\textwidth}
        \centering
        \includegraphics[width=\textwidth]{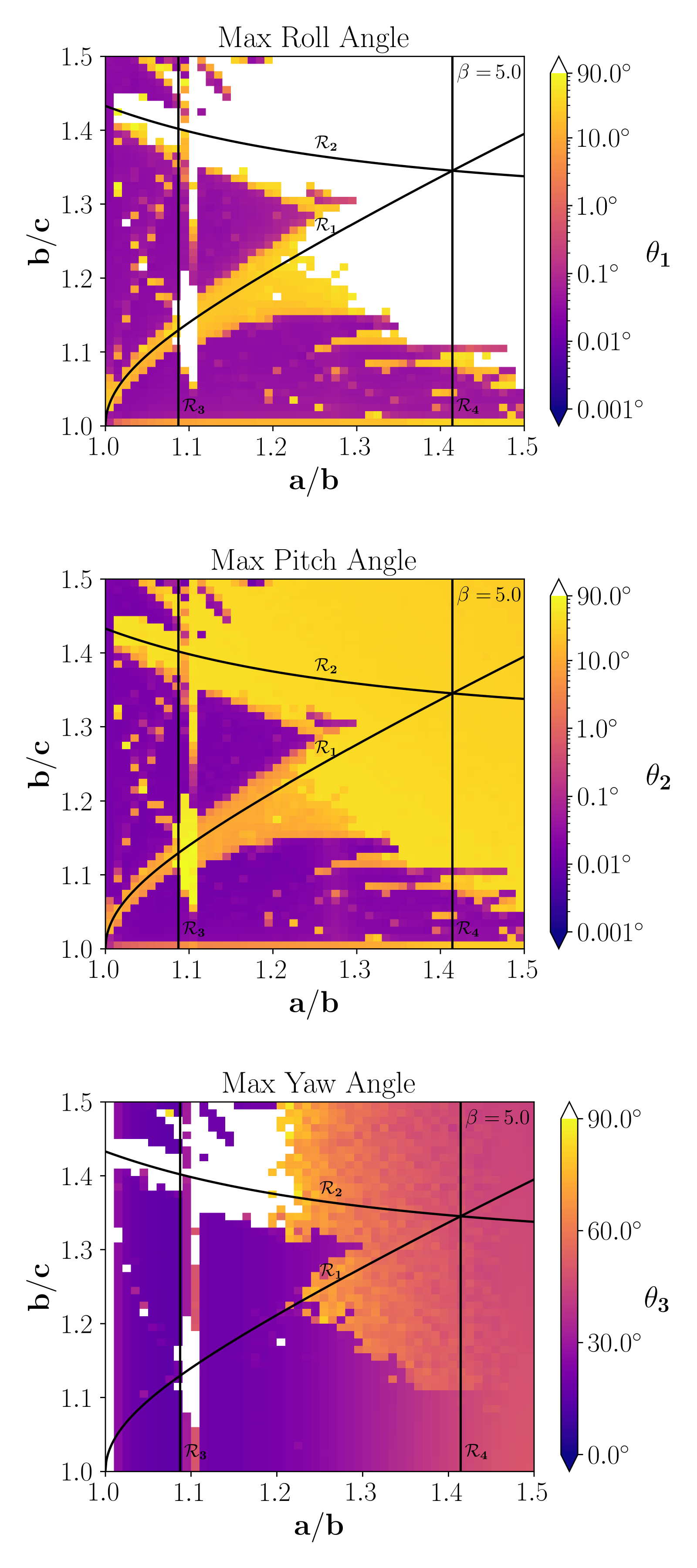} 
        \caption{Max Euler angles, $\beta=5$.}
        \label{fig: gubasEuler_beta5}
    \end{subfigure}
    \caption{The maximum Euler angles over a one-year simulation for $\beta=1, 3$ and $5$. Cases that exceed $90^{\circ}$ are shown in white. The pitch angle ($\theta_{2}$) is defined between $0^{\circ}$ and $90^{\circ}$, and therefore never exceeds $90^{\circ}$. The uncoupled, analytic resonance locations from Table \ref{table: resonances} are overlaid. Over half of the parameter space is attitude unstable, dominated by rolling about the secondary's long axis, when $\beta=5$.}
    \label{fig: gubasEuler_beta135}
\end{figure}

The attitude stability over the full parameter space when $\beta=1$ is shown on Fig.\ \ref{fig: gubasEuler_beta1}, with the maximum roll, pitch, and yaw angles achieved by Dimorphos over the full simulation. When Dimorphos's attitude is broken into its constituent components, we can see very clearly that the same resonances found with the simplified 3D model are driving the instabilities in the fully coupled problem. The most significant result here is the tendency for more-elongated secondaries to roll about their long axis, reaching angles of nearly $90^{\circ}$ (almost rolling over), an effect that would have gone unnoticed by just computing the libration angle. 

Figure \ref{fig: gubasEuler_beta3} shows the maximum Euler angles when $\beta=3$. Note the difference in colorbar scale for the yaw angle between Figs.\ \ref{fig: gubasEuler_beta1} and \ref{fig: gubasEuler_beta3}. When $\beta$ is increased to 3, we see the overall shape of the instability region stay the same, with the size of the region and magnitude of the maximum angle increasing. The most obvious difference is that many of the more-elongated secondaries exceed a roll angle of $90^{\circ}$, indicating that they have either rolled over or are continuously rolling. In most of these cases, the secondary has rolled over without technically breaking from synchronous rotation (defined by the libration angle exceeding $90^{\circ}$). The implications for this are discussed in Section \ref{subsection: barrel instability}. We also see that a select few cases have exceeded $90^{\circ}$ in yaw near the intersection of resonances.


\begin{figure}[t!]
    \centering
    \begin{subfigure}[t]{0.5\textwidth}
        \centering
        \includegraphics[width=\textwidth]{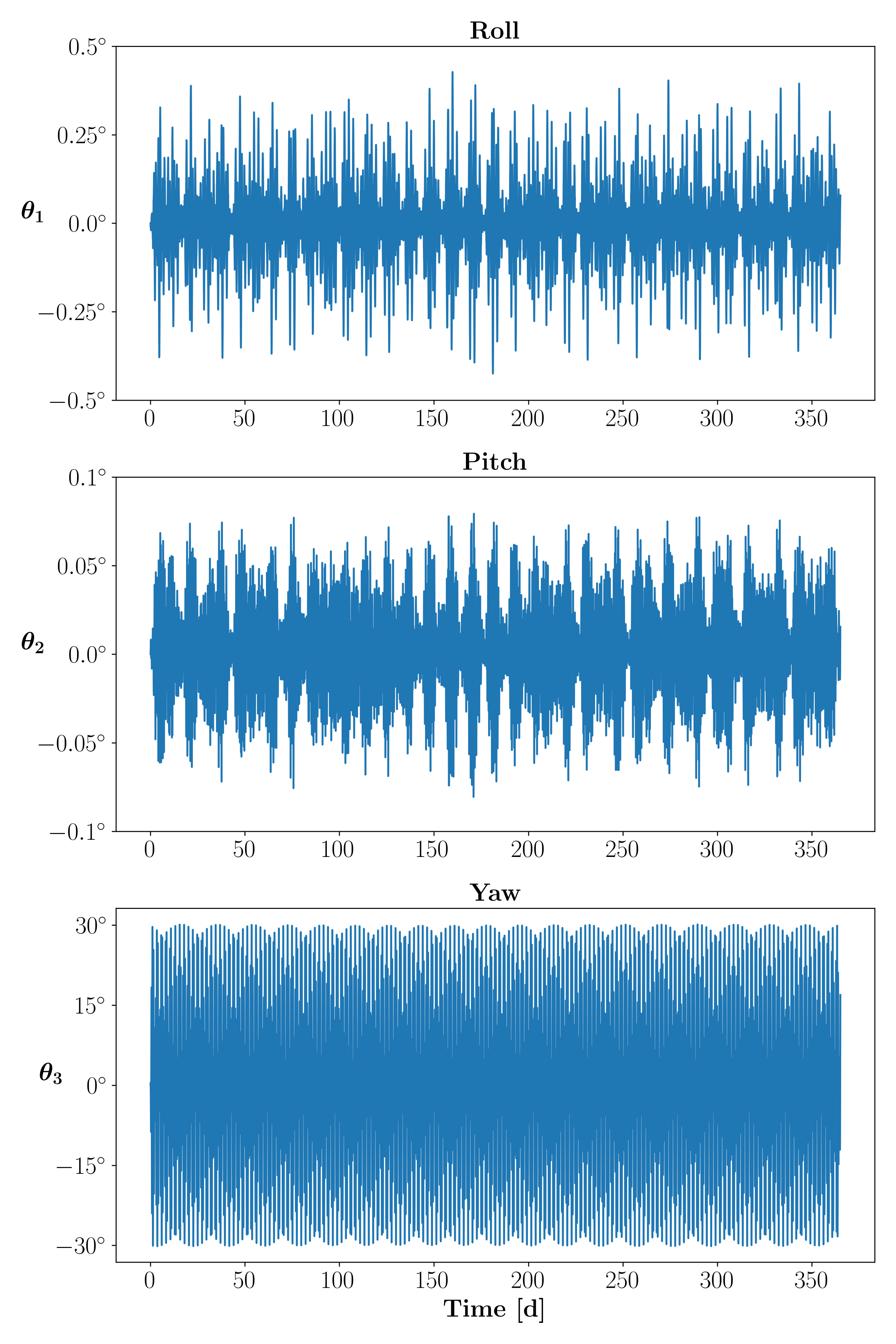}
        \caption{\centering $\beta=5$, $a/b=1.29$, $b/c=1.30$\newline $a =105.40597\text{ m}$\newline $b = 81.71005\text{ m}$\newline $c = 62.85389\text{ m}\newline$}
        \label{fig: eulerAngles_ab1.29bc1.30}
    \end{subfigure}%
    ~ 
    \begin{subfigure}[t]{0.5\textwidth}
        \centering
        \includegraphics[width=\textwidth]{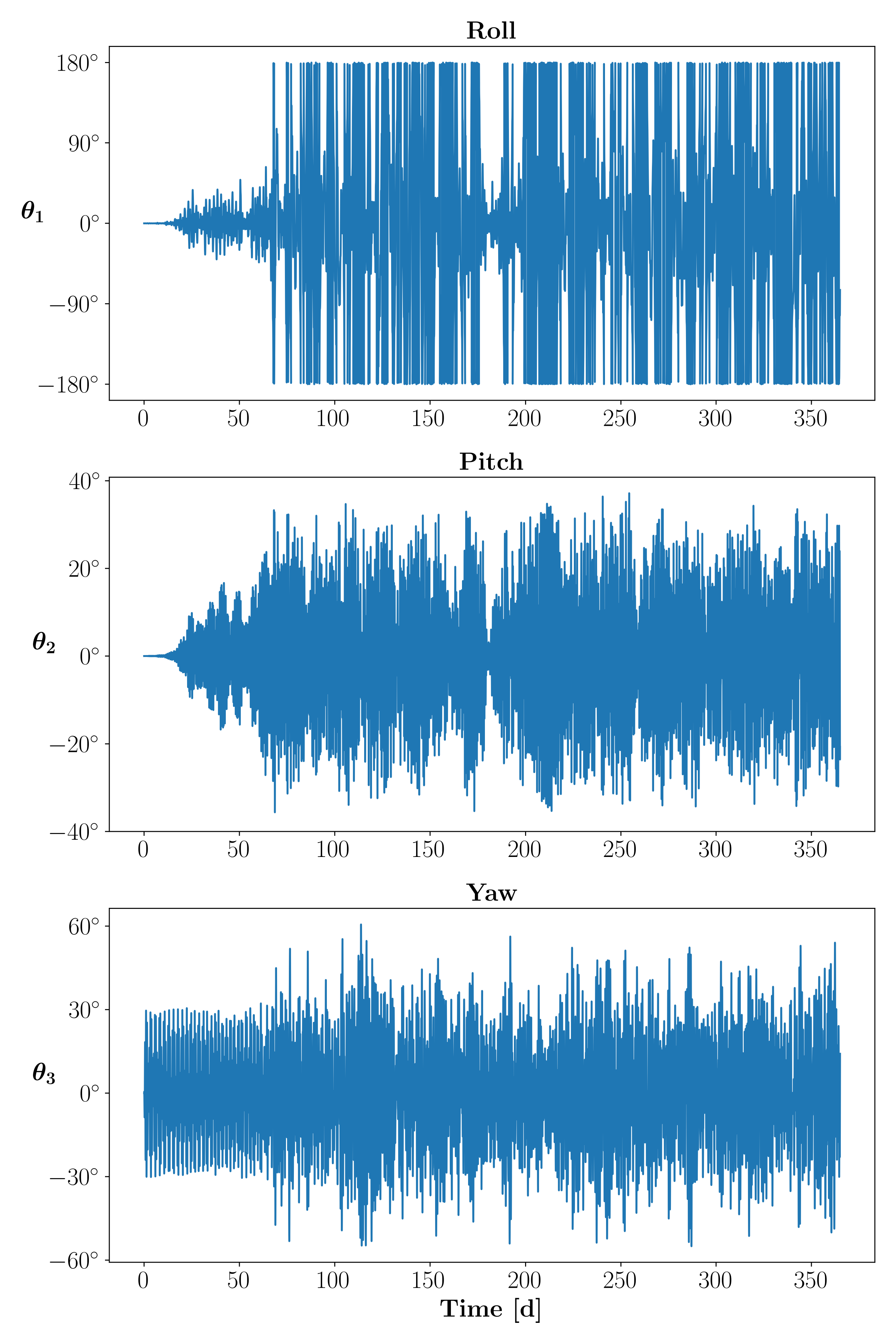}
        \caption{\centering $\beta=5$, $a/b=1.29$, $b/c=1.31$\newline $a =105.67555\text{ m}$\newline $b = 81.91903\text{ m}$\newline $c = 62.53361\text{ m}$\newline}
        \label{fig: eulerAngles_ab1.29bc1.31}
    \end{subfigure}
    \caption{Time-series plots of the secondary's three Euler angles for two different secondary shapes: $a/b=1.29, b/c=1.30$ (a stable case) and $a/b=1.29, b/c=1.31$ (an unstable case). The physical semi-axis lengths are listed in the subcaptions. Although these two shapes differ by a fraction of a meter in their respective semi-axes lengths, their attitude evolutions vary drastically. Note the different angle-axis scales on each plot. Again, the DART impact occurs at time $=0$.}
    \label{fig: eulerAngles_beta5}
\end{figure}

Finally, for completeness, we show the same maximum Euler angle plots when $\beta=5$ in Fig.\ \ref{fig: gubasEuler_beta5}, where over half of the parameter space has become unstable and exhibits chaotic tumbling motion. To get a rough understanding of what this tumbling motion looks like over time, we show time-series plots of the three Euler angles for two shapes of the secondary on Fig.\ \ref{fig: eulerAngles_beta5}. Fig.\ \ref{fig: eulerAngles_ab1.29bc1.30} corresponds to $a/b=1.29$ and $b/c=1.30$, which lies in the ``stable island'' near the middle of the parameter space from Fig.\ \ref{fig: gubasEuler_beta5}. This subplot shows regular, stable motion over the course of the entire year. Shown in Fig.\ref{fig: eulerAngles_ab1.29bc1.31}, the attitude instability sets in after ${\sim}15$ days, with long-axis rolling commencing at ${\sim}60$ days when the $b/c$ axis ratio is changed only slightly from $1.30$ to $1.31$. This highlights how sensitive the instability can be to the shape of the secondary. To put this in perspective, the physical extents along each axis of these two body shapes differ by a fraction of a meter compared to the ${\sim}100\text{ m}$ scale of the full axes, yet one enters a chaotic spin state and the other remains stable.


\section{Discussion}
\label{section: discussion}

\subsection{Implications of a post-impact tumbling state}
\label{subsection: implications}

One concern about a post-impact tumbling state is that the periodic (and chaotic) exchange of angular momentum between the mutual orbit and Dimorphos's spin state could affect the post-impact ground-based measurements of the orbit period. This could cause a portion of the orbit period change to be misattributed to the DART impact, and complicate the estimate of $\beta$ based on the orbit period change, which is a Level 1 mission requirement. To get a rough idea of how important this effect might be, we compare the energy of the mutual orbit (which determines the orbit period) with Dimorphos's spin energy. To first order, we can estimate the energy in the mutual orbit by assuming the bodies behave as point-masses, so the problem reduces to a Keplerian orbit. Then the orbital energy can be written as $E_{\text{orb}} \simeq -\frac{GM_{\text{A}}M_{\text{B}}}{2r}$, where $M_{\text{A}}$ and $M_{\text{B}}$ are the respective body masses, $r$ is the binary semimajor axis, and $G$ is the standard gravitational parameter. The secondary's spin energy can be written as $E_{\text{B}} = \frac{1}{2}C\omega^{2}$, where $C$ is Dimorphos's largest principal moment of inertia, and $\omega$ is the spin rate. If Dimorphos is a uniform ellipsoid in synchronous rotation, we can write $E_{B} = \frac{1}{2}\frac{M_{B}}{5}(a^{2}+b^{2})\big(\frac{2\pi}{P_{\text{orb}}}\big)^{2}$, where $a$ and $b$ are Dimorphos's semimajor and semi-intermediate axis lengths, and $P_{\text{orb}}$ is the binary orbit period. Choosing axial ratios for Dimorphos that lie in the middle of the parameter space studied in this work ($a/b=b/c=1.25$) along with nominal parameters for the system,\footnote{Based on current observations and an assumed Keplerian orbit, we have $M_{\text{sys}}= 5.37\times10^{11}\text{ kg}$, $M_{\text{B}} = 4.97\times10^{9} \text{ kg} $, $r = 1190 \text{ m} $, $P_{\text{orb}}=11.9217\text{ h}$. Further, if $a/b=b/c=1.25$, we have $a=101.875\text{ m}$ and $b=81.5 \text{ m}$.} we find a value of $E_{B}/E_{\text{orb}}\simeq0.25\%$\footnote{A similar approach can be used for the spin and orbital angular momenta, where we get $L_{B}/L_{\text{orb}}\simeq2\times10^{-5}$.}, meaning that the energy in Dimorphos's rotation is much less than the energy of the mutual orbit. This indicates that even relatively large changes in Dimorphos's spin state should be small compared to the energy in the mutual orbit, meaning that any spin-state-induced changes to the orbit period should be a small effect. In other words, spin-orbit coupling or a chaotic tumbling state is unlikely to affect the Level 1 requirement to measure $\beta$. The uncertainties in the $\beta$ measurement will be dominated by uncertainties in Dimorphos's mass, the impact location and surface normal, and the ejecta cone geometry {\color{red}\hypersetup{citecolor=red}\citep{rivkin2021}\hypersetup{citecolor=DarkSlateGrey}}. However, a focused study on secular changes to the mutual orbit resulting from a tumbling secondary could be valuable. This effort is planned for future work, as this study concentrated only on Dimorphos's spin state resulting from the DART impact.

If Dimorphos begins chaotically tumbling, it is possible but highly unlikely to be detected via ground-based observations, given the required precision to measure sufficiently small fluctuations in the lightcurve (Pravec, P., personal communication, 2021). Of course, this is dependent on the details of Dimorphos's shape, the observing geometry, and available telescope facilities, but even under ideal conditions, such a measurement is likely unachievable from the ground during the first few years after the impact. However, if the tumbling persists for several years, Hera will be able to provide detailed measurements upon arrival \citep{michel2018}. In particular, the Hera mission design requirements state that the spin pole orientation shall be measured to a precision of 1$^{\circ}$, and therefore, Hera will be able to accurately characterize any deviation larger than that. The visible and infrared cameras as well as the radio science on-board Hera can all be used to constrain Dimorphos's spin state, provided sufficient accuracy in the orientation and position of the spacecraft. Hera's two CubeSats, especially once landed on the surface, will increase the measurement precision of Dimorphos's spin state. The synergies between the instruments of Hera will thus offer the opportunity to investigate the rotational state of Dimorphos, checking the predictions and consequences of the DART impact. We note that a chaotic tumbling state should not influence most of Hera operations. However, if Dimorphos is tumbling, the attitude reconstruction of Hera will have to rely on star tracking until the rotation state is constrained well enough to rely on landmarks. Only during very close proximity operations, like close flybys, will Dimorphos's spin state need to be taken into account. This could potentially add challenges for close operations with the CubeSats and their landing trajectories.

\subsection{The Barrel Instability}
\label{subsection: barrel instability}

The ``barrel instability'' is a low-energy instability in which an elongated secondary in an eccentric orbit enters a long-axis rotation state while remaining tidally locked \citep{cuk2020_dda}. We find similar behavior in the simulations presented here (indicated by large roll angles), suggesting that a post-impact rolling state about the long axis is a possible outcome. The GUBAS simulations show that the secondary can rotate about its long axis, even if the secondary's long axis remains aligned with the line of centers (i.e., libration angle $<90^{\circ}$).

Through the YORP effect, the spins of asteroids can be altered by the absorption of sunlight and anisotropic re-emission as thermal radiation resulting in a net torque \citep{rubincam2000}. This process has an analog for binary systems, commonly referred to as the binary-YORP or BYORP effect, whereby a tidally-locked satellite's thermal emission contributes a net torque to the mutual orbit, leading to secular changes in the mutual orbit's semimajor axis and eccentricity \citep{cuk2005}. Because the BYORP effect requires a synchronous secondary, a rolling or tumbling state resulting from a DART-like impact would terminate the BYORP process, even if the secondary's long axis remains aligned with the LOC.

The orbital solution for the Didymos binary is achieved by fitting the timing of mutual events to an orbit model \citep{scheirich2009}. The model includes the quadratic drift of the secondary's mean anomaly due to secular changes in the semimajor axis that result from the combined effect of BYORP and tides. Based on historical data and recent observations during the 2020--2021 apparition, the latest best-fit solution to Dimorphos's mean anomaly acceleration is $0.13 \pm 0.14\text{ deg yr}^{-2}$ (Scheirich and Pravec, personal communication, 2021). The mean anomaly acceleration corresponds to a drift in the mean motion of $\dot{n} = 4.45\pm4.91\times10^{-18}\text{ rad s}^{-2}$ and a drift in the semimajor axis of $\dot{a} \simeq -0.076\text{ cm yr}^{-1}$. The measured mean anomaly acceleration is extremely small, with the 3-sigma uncertainty including zero, meaning that any secular changes to the orbit are small, if not zero. This implies that Dimorphos may be in (or very close to) a BYORP-tide equilibrium, a state predicted analytically by \cite{jacobson2011b} in which tides and BYORP effectively cancel each other. If the DART impact excites an attitude instability, the resulting rolling and/or tumbling motion of Dimorphos would shut off the BYORP process \citep{cuk2020_dda}. Because secular changes in the semimajor axis due to tides do not require a synchronous secondary\footnote{Although spin synchronization of the secondary is the fastest tidal process, it is not a prerequisite for the mutual orbit to expand through tides. So the mutual orbit would evolve concurrently as the secondary begins to re-synchronize.} \citep{goldreich2009, MurrayDermott}, the termination of BYORP would allow the mutual orbit to begin to evolve solely under the influence of tides. It should be noted here that we do not know how effective the tides are at suppressing Dimorphos's attitude instability. It may be possible, but unlikely, that tides could shrink the parameter space of cases that become attitude unstable to begin with. Or, tides may suppress the attitude instability after it has begun, so that the system finds the BYORP-tide equilibrium again. In either case, this knowledge could be used to place a constraint on the tidal parameters of the system.

If, however, the BYORP effect is terminated due to an attitude instability, it could present a unique opportunity to constrain the tidal parameters of the system. Under the assumption that Didymos is currently in a BYORP-tide equilibrium, the quantity $BQ/k$ can be estimated from observable quantities, where $B$ is the BYORP coefficient, $Q$ is the tidal quality factor, and $k$ is the tidal Love number. The BYORP coefficient, $B$, is a unitless parameter that depends only on the shape of the secondary and quantifies the degree of symmetry of the body. The quality factor $Q$ describes the efficiency of energy dissipation through tides \citep{goldreich1966}. Finally, the Love number $k$ describes the gravitational response of a body to tides and can be thought of as the ratio of additional gravitational potential produced by a body in response to the perturbing potential to the perturbing potential itself. \cite{jacobson2011b} derive an equation for $BQ/k_{p}$ for a binary asteroid in BYORP-tide equilibrium,

\begin{equation}
    \frac{BQ}{k_{\text{p}}} = \frac{2\pi\omega^{2}_{\text{d}}\rho R_{\text{p}}^{2}q^{4/3}}{H_{\odot}(a/R_{\text{p}})^{7}},
\end{equation}
where $k_{\text{p}}$ is the Love number of the primary,  $\omega_{\text{d}}$ is the spin-disruption limit, $\rho$ is the bulk density of both bodies, $R_{\text{p}}$ is the radius of the primary, $q$ is the secondary-to-primary mass ratio, $H_{\odot}$ is the radiation pressure, and $a$ is the binary semimajor axis. The spin disruption limit can be written as $\omega_{\text{d}} = \sqrt{4\pi G \rho/3}$ and we assume that both the primary and secondary have the same bulk density, so there is only one value for $\rho$. We can also write $H_{\odot} = F_{\odot}/\big(a_{\odot}^{2}\sqrt{1-e_{\odot}^{2}}\Big)$, where $a_{\odot}$ and $e_{\odot}$ are the heliocentric semimajor axis and eccentricity, and $F_{\odot}$ is the solar radiation constant, which is ${\sim}10^{22}\text{ g cm s}^{-2}$ \citep{mcmahon2010}.

Based on nominal parameters for Didymos from the literature, we have $\rho=2.17 \text{g cm}^{-3}$, $R_{\text{p}} = 390\text{ m}$, $q = 0.00926$, $a_{\odot} = 1.644 \text{ au}$, and $e_{\odot} = 0.384$ \citep{naidu2020, scheirich2009, pravec2006}\footnote{These values are derived or taken from the listed references. Note that they may be slightly different than the values used in the simulations presented in this paper}. Plugging in numbers, we find $BQ/k_{\text{p}} \simeq 555$. A proper treatment of error propagation would show that the uncertainties on $BQ/k$ can be relatively large \citep{jacobson2011b}, due to the uncertainties in all the observable parameters. However, these will decrease with future ground-based observations and DART imagery leading up to the moment of impact. 

If the secondary then enters a tumbling state following the DART impact, BYORP will shut off and the system will evolve primarily through tides. After such an excitation, the tides between the binary components will dissipate energy working to bring it back to a minimum energy state. Spin synchronization, the fastest-evolving tidal process, will be driven by tides raised on the secondary by the primary. Meanwhile, tides raised on the primary by the secondary will act to increase the mutual semi-major axis and eccentricity. However, tides raised on the secondary will act to damp the eccentricity along with the radial and librational tides\citep{goldreich2009, MurrayDermott}. The ``barrel instability'', or rolling about the secondary's long axis will not be damped by any of these tides. However, it is sensitive to obliquity tides so long as the secondary's spin axis is unaligned with the mutual orbit pole. Given the limited knowledge of $Q/k$ for asteroids and the relative strengths of each tidal mechanism, we provide a simple order-of-magnitude estimate for only the rate of change in the semimajor axis due to the tides raised on the primary by the secondary. If the tides are strong enough, then it may be possible to measure a secular change in the mean anomaly (and therefore the semimajor axis), either through ground-based observations or when Hera makes its rendezvous in 2026. If a change in the semimajor axis can be measured $\big(\frac{da}{dt}\big)$, then it would be possible to estimate $Q/k$ to constrain the tidal properties of the system. For low eccentricity, the rate of change in the semimajor axis can be approximated as \citep{goldreich2009},

\begin{equation}
    \label{eq: tidalEvolution}
    \frac{1}{a}\frac{da}{dt} = 3qn\frac{k_{\text{p}}}{Q}\bigg(\frac{R_{\text{p}}}{a}\bigg)^{5},
\end{equation}
where $n$ is the mean motion. Although the ratio $k/Q$ (or $Q/k$) is commonly used to parameterize a body's tidal response, its value and scaling relationship with other physical properties remain open questions for rubble piles. For example, \cite{goldreich2009} find that $k$ should scale linearly with the body's radius, while \cite{jacobson2011b} find that the scaling should vary inversely with the radius. For a Didymos-like system, \cite{nimmo2019} suggest that $Q/k\sim10^{5}$. For the sake of demonstration, we can plug this value into Eq.\ \eqref{eq: tidalEvolution} to get an estimate of $\frac{da}{dt}\simeq 0.58 \text{ cm yr}^{-1}$. Although this is a small number, a value of ${\sim}1\text{ cm s}^{-1}$ has been measured for near-Earth binary Moshup through ground-based measurements of the system's mean anomaly acceleration \citep{scheirich2021}. In addition, the rendezvous of the Hera spacecraft will make it possible to make much-higher-precision measurements than are possible from the ground. Furthermore, spin synchronization is the fastest-evolving tidal process, meaning that Hera should be able to measure a secular change to Dimorphos's spin if its semimajor axis drift is also measurable. 

The BYORP coefficient, $B$, depends only on the secondary's shape, and can therefore be calculated from a shape model which will become available after the DART impact (and refined with Hera measurements). \cite{scheirich2015} show that an independent measurement of $B$ derived from a shape model can be combined with an independent measurement of $BQ/k$ at the BYORP-tide equilibrium to obtain a non-degenerate solution for both $B$ and $Q/k$. When a shape model for Dimorphos is created, it will be possible to do this, so long as the system is actually in the BYORP-tide equilibrium prior to impact. Although this approach will be dependent on the resolution scale of the shape model and further complicated by the fact that the crater formed by DART will slightly alter Dimorphos's shape and affect $B$. If the barrel instability is triggered following the impact, a separate, independent measurement of $Q/k$ can be made using Eq.\ \eqref{eq: tidalEvolution}. Therefore it may be possible to have three independent measurements constraining the BYORP and tidal evolution of the system: $BQ/k$ prior to impact, $Q/k$ after impact, and $B$ from the secondary's shape model.

Such a measurement of the tidal and BYORP parameters of the system is of course complicated by a variety of factors. First, an excitation of the secondary's spin state could induce local slope failure, leading to possible deformation and surface motion. This would create a direct dynamical effect on the system's evolution if the secondary's shape changes plus additional energy dissipation due to surface motion that would muddle any measurement of tidal dissipation. Second, the crater created by the DART impact could have a minor effect on the secondary's mass distribution and potentially a major effect on its BYORP coefficient, given the high sensitivity of $B$ to the body shape. Third, the primary's spin rate sits right at the critical spin limit for a non-cohesive spherical body, meaning that the acceleration required to loft material near the surface is quite small. There is a possibility that "tidal saltation", a process by which the tidal acceleration of the secondary lofts material off the surface of the primary \citep{harris2009}. This results in a transfer of angular momentum from the primary to the mutual orbit as the material is lofted and falls back to the surface. This process would enhance the effect of tidal evolution in addition to serve as a 'breaking' mechanism that prevents YORP spin-up from driving the primary beyond its critical spin limit. A related effect could be larger global-scale reshaping of Didymos and its effect on the mutual orbit \citep{hirabayashi2017, hirabayashi2019}. These examples are only a select few effects that could lead to long-term changes to Dimorphos's mutual orbit and spin state that could complicate any future measurements of the system's tidal parameters.

\section{Conclusions}
We presented three independent methods---one analytic and two numerical---to study the attitude dynamics of Dimorphos. The analytic model found four fundamental periods of motion relating to the mean motion and the free libration, precession, and nutation frequencies of the secondary. At the resonance locations among these various frequencies, we predicted that unstable motion could be possible. Then, using the ``simplified 3D model'', we simulated the post-impact attitude evolution of the secondary when $\beta=1$ and $3$, where we found several of the predicted resonances and verified that they did produce unstable motion. Using fast Lyapunov indicators, we demonstrated that the secondary's attitude evolves chaotically at the resonance locations. Then, using GUBAS, we verified the results of the simplified 3D model. We found that the simplified 3D model predicted the instability regions correctly but overestimated the magnitude in the libration amplitudes, but GUBAS (or any other fully-coupled F2BP code) is necessary to accurately predict the amplitude of oscillations in the instability regions. The GUBAS simulations also demonstrated that the secondary is especially prone to unstable rotation about its long axis. The implications for a post-impact tumbling and/or rolling state were discussed, including the consequences for Hera, the possibility of terminating any BYORP drift, and measuring the tidal parameters of the system.

In this work, we assumed an idealized impact in which the DART momentum is transferred entirely within the mutual orbit plane with no instantaneous change to the secondary's spin (i.e., a centered impact). In reality, the DART momentum vector will nominally be misaligned with respect to the mutual orbit plane by an angle between $5^{\circ}$ and $30^{\circ}$ with respect to the orbit plane, imparting some nonplanar motion to the mutual orbit. In addition, the DART impact is unlikely to be perfectly aligned with the center of mass, and will deliver an instantaneous torque to the secondary. We expect that these effects will lead to the development of attitude instabilities at earlier times and make a larger portion of the parameter space unstable. In addition, the pre-impact state of the system was in an idealized relaxed dynamical state. For these reasons, the results presented in this work should be viewed as a \textit{lower bound} on Dimorphos's post-impact spin dynamics. The exploration of non-planar, off-center impacts, including non-relaxed pre-impact states is planned for future work.   

Finally, with ongoing GUBAS code development, including a tidal evolution model, we plan to explore the role that tides may play in the Didymos system. As a function of the tidal parameters $k$ and $Q$, we will investigate the likelihood of exciting the attitude instabilities demonstrated in this work, and predict the dissipation of Dimorphos's spin state, if excited. This may make it possible to predict the binary orbital evolution on timescales relevant to the Hera mission.

\section*{Acknowledgements}

H.F.A. would like to thank Douglas Hamilton, Kevin Walsh, and Jean-Baptiste Vincent for useful discussions.

This study was supported in part by the DART mission, NASA Contract \#NNN06AA01C to JHU/APL. M.C. acknowledges support from the NASA Solar System Workings program (80NSSC21K0145). I.G., K.T., P.M., and Y.Z. acknowledge funding support from ESA and the European Union’s Horizon 2020 research and innovation programme under grant agreement No. 870377 (project NEO-MAPP). O.K. acknowledges funding support from the NEO-MAPP project, the Belgian Science Policy Office (BELSPO) through the ESA/PRODEX Program, and the European Union's Horizon 2020 research and innovation program PIONEERS-project under grant agreement number 821881. P.M. and Y.Z. also acknowledge funding support from CNES. Y.Z. acknowledges funding from the Universit\'{e} C\^{o}te d'Azur ``Individual grants for young researchers'' program of IDEX JEDI. A small portion of this research was carried out at the Jet Propulsion Laboratory, California Institute of Technology, under a contract with the National Aeronautics and Space Administration (80NM0018D0004).

Some of the simulations herein were carried out on The University of Maryland Astronomy Department's YORP cluster, administered by the Center for Theory and Computation.

\bibliographystyle{cas-model2-names}

\bibliography{refs}

\appendix
\section{The simplified 3D model}
The ``simplified'' 3D model consists of the planar mutual orbit described in Appendix \ref{appendix: J2+ell} combined with the 3D attitude of the secondary described in Appendix \ref{appendix: 3D attitude}. This approach has the advantage of being an extremely fast approach to numerically model the system dynamics. However, it should be noted that the mutual orbit and secondary spin state are not fully coupled. The mutual separation and velocity is predetermined and set by the equations of motion in Appendix \ref{appendix: J2+ell}, and the secondary's spin and attitude are then propagated based on equations \eqref{eq: attitude_EOM_1}--\eqref{eq: attitude_EOM_3}.

\begin{figure}
\centering
\includegraphics[width=0.5\textwidth]{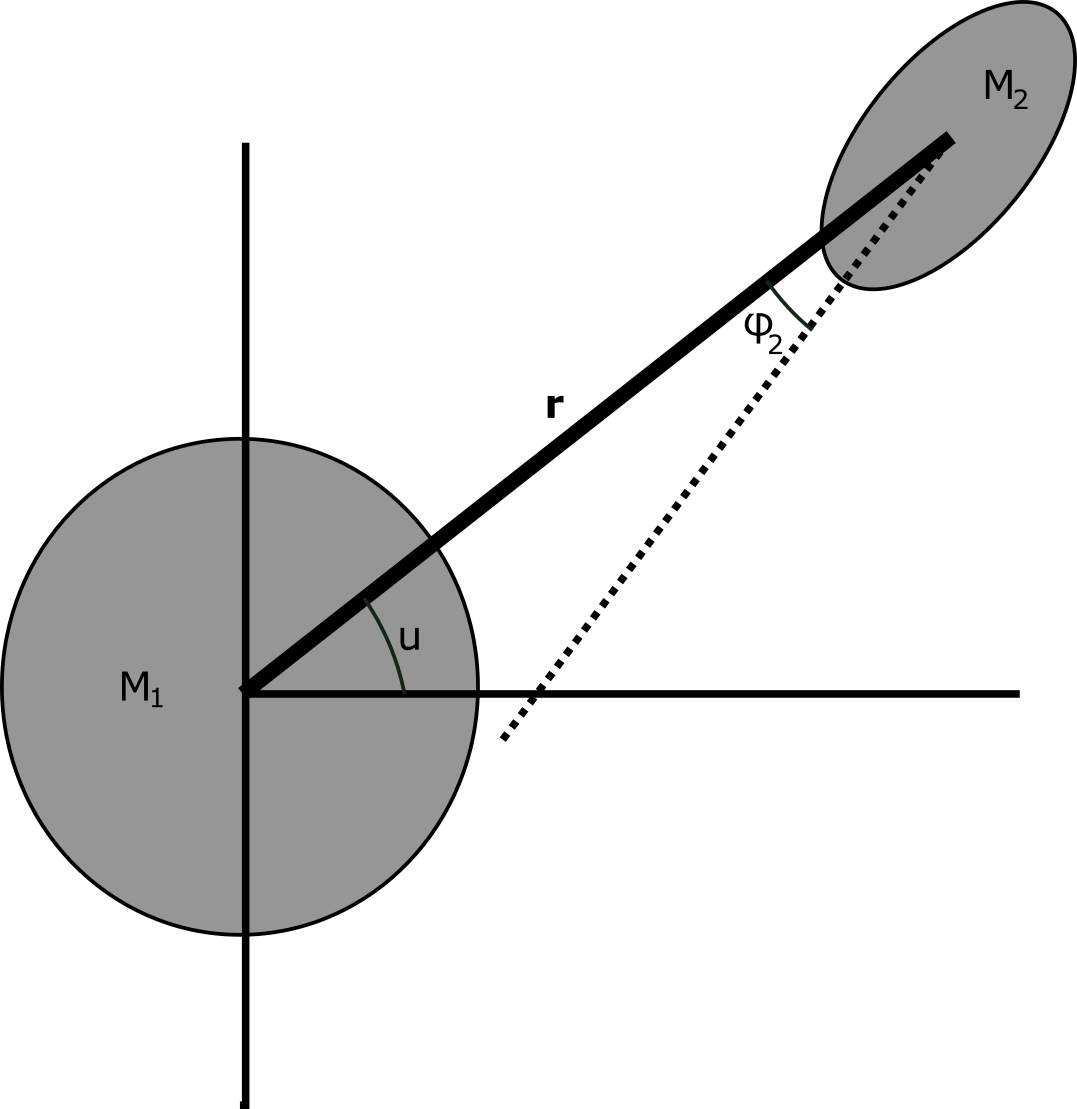}
\caption{The simplified configuration of the planar $J_2$-ellipsoid model. \label{fig: J2ell}}
\end{figure}

\subsection{The planar``$J_{2}+$ellipsoid'' model}
\label{appendix: J2+ell}

The planar ``$J_{2}+$ellipsoid'' model consists of two extended bodies, with masses $M_1$ and $M_2$, that orbit and interact with each other via their mutual gravitational potential, as shown in Fig.\ \ref{fig: J2ell}. The primary is assumed to be an oblate spheroid with moments of inertia $A_1 = B_1 < C_1$, while the secondary is a triaxial ellipsoid with moments of inertia $A_2 < B_2 < C_2$. The Hamiltonian of the system reads,

\begin{equation}
\mathcal{H} = \frac{1}{2} \left( \frac{P_{\phi_2}}{C_2} + \frac{(P_u - P_{\phi_2})^2}{r^2 \mu} + \frac{P_r^2}{\mu} \right) + V(r,\phi_2),
\end{equation}
where $r$ is the distance between the two mass centers and $\phi_2$ is the angle between the secondary's long axis and the radius vector $\bf r$. The angle $\phi_2$ is also known as the libration angle of the secondary. As shown in Fig.\ \ref{fig: J2ell}, $u$ is the angle of the radius vector with respect to an inertial system of reference. Then, $P_r$, $P_{\phi_2}$, and $P_u$ are the conjugate momenta and $\mu = M_1 M_2/( M_1+M_2)$. 

The mutual potential, up to second-order expansion, is \citep{McMahon2013},
\begin{align}
V(r,\phi_2) =& -\frac{G M_1 M_2}{r} - \frac{G M_2}{2 r^3} (2 A_1+C_1) - \frac{G M_1}{2 r^3} (A_2+B_2+C_2) \nonumber \\
&+ \frac{3 G M_2 A_1}{2 r^3} + \frac{3 G M_1}{4 r^3} (A_2+B_2 - (B_2-A_2) \cos(2 \phi_2)),
\end{align}  
where $G$ is the gravitational constant. 


The angle $u$ is ignorable and therefore the momentum $P_u$ is a constant of the motion. For a fixed separation $r_0$ there exists an equilibrium solution, corresponding to a circular orbit with orbital frequency $\dot{u}_{eq}$  \citep{McMahon2013},
\begin{equation}
\dot{u}_{eq} = \sqrt{\frac{G (M_1+M_2)}{r_0^3} \left(1 + \frac{3}{2 r_0^2} \left( \frac{C_1-A_1}{M_1}+\frac{B_2+C_2-2 A_2}{M_2} \right) \right) },
\label{eq:equilib} 
\end{equation} 
and the conjugate conserved momentum is,
\begin{equation}
P_u = ( C_2 +  r_0^2 \mu) \dot{u}_{eq}.
\end{equation}

Notice that if both the binary separation $r$ and the orbital frequency are assumed to be known quantities, Eq.\ \eqref{eq:equilib} could be solved to provide the total mass of the system $(M_1+M_2)$. This strategy was followed to obtain the relaxed equilibrium conditions assuming a constant-density ellipsoid with axial ratios $a/b$ and $b/c$ for Dimorphos. 

\subsection{The 3-dimensional rotation of the secondary}
\label{appendix: 3D attitude}
In order to study the attitude stability of Dimorphos with the simplified 3D model, we employ the 3-1-2 set of Euler angles\footnote{The simplified 3D model uses the 3-1-2 Euler angle set in its numerical integrations. However, the Euler angle plots in the manuscript use the 1-2-3 Euler angle set.} $(\theta,\phi,\psi)$. Assuming a planar orbit for Dimorphos described by $r(t)$ and $u(t)$, Euler's rigid body equations read \citep{Wisdom1984},
\begin{align}
A_2 \dot{\omega}_x - \omega_y \omega_z (B_2 - C_2)  = - \frac{3 G M_1}{r^3} \beta \gamma (B_2-C_2), \\
B_2 \dot{\omega}_y - \omega_z \omega_x (C_2 - A_2)  = - \frac{3 G M_1}{r^3} \alpha \gamma (C_2-A_2), \\
C_2 \dot{\omega}_z - \omega_x \omega_y (A_2 - B_2) = - \frac{3 G M_1}{r^3} \alpha \beta (A_2-B_2), 
\end{align}
where $\omega_x$, $\omega_y$ and $\omega_z$ are the rotational angular velocity components with respect to the Dimorphos body-fixed axes $x$, $y$ and $z$ respectively. The direction cosines $\alpha$, $\beta$ and $\gamma$ are given from the relations,
\begin{align}
\alpha &= \cos{\psi} \cos{(\theta-u)} - \sin{\psi} \sin{\phi} \sin{(\theta-u)}, \\
\beta & = - \cos{\phi} \sin{(\theta-u)}, \\
\gamma & = \sin{\psi} \cos{(\theta - u)} + \cos{\psi} \sin{\phi} \sin{(\theta-u)}.
\end{align}

Finally, the corresponding kinematic equations for the specific set of Euler angles read,
\begin{align}
\label{eq: attitude_EOM_1}
\dot{\theta} &= \sec{\phi} (\omega_z \cos{\psi} - \omega_x \sin{\psi}), \\
\label{eq: attitude_EOM_2}
\dot{\phi} &= \omega_x \cos{\psi} + \omega_z \sin{\psi},\\
\label{eq: attitude_EOM_3}
\dot{\psi} &= \omega_y - \omega_z \cos{\psi} \tan{\phi} + \omega_x \sin{\psi} \tan{\phi}. 
\end{align}

The planar solution for $r(t)$ and $u(t)$ is inserted into Euler's equations of motion and the 3-dimensional attitude dynamics are propagated. Given no out-of-plane excitation (i.e., $\phi = \psi = \omega_x = \omega_y = 0$), then the computed librational solution is equivalent to the planar one, 
$$
\phi_{2,3D} = u(t) - \theta(t) = \phi_2(t).
$$

If a small excitation is assumed in the out-of-plane rotation (i.e., $\omega_x = \omega_y = 10^{-15}$), then for some initial conditions $\phi_{2,3D} \approx \phi_2$ while in other cases the two solutions diverge. In order to study these instabilities in more detail, an analysis based on the linearized system of equations is required.

First the set of Euler's equations is cast into a Hamiltonian form via the Legendre transformation,
\begin{align}
P_\theta &= - A_2 \omega_x \cos{\phi} \sin{\psi} + B_2 \omega_y \sin{\phi} + C_2 \omega_z \cos\phi \cos\psi, \\
P_\phi &= A_2 \omega_x \cos\psi + C\omega_z \sin\psi, \\
P_\psi &= B_2 \omega_y,
\end{align}

and substituting $\omega_x$, $\omega_y$, and $\omega_z$ in the Hamiltonian of the rotation, we have,
\begin{equation}
\mathcal{H}_{rot} = \frac{1}{2} (A_2 \omega_x^2 + B_2 \omega_y^2+ C_2 \omega_z^2) + V(r,u,\theta,\phi,\psi).
\end{equation}
with,
\begin{equation}
V(r,u,\theta,\phi,\psi) = \frac{3 G M_1}{2 r^5} \mathbf{r} \mathcal{A}_2^{T} I_2 \mathcal{A}_{2} \mathbf{r},
\end{equation}
where $I_2 = \text{diag}(A_2,B_2,C_2)$, $\mathbf{r} = (r \cos u, r \sin u, 0)$ and $\mathcal{A}_{2} = R_y(\psi) R_x(\phi) R_z(\theta)$. 


An equilibrium solution of the planar $J_2$-ellipsoid Hamiltonian $\mathcal{H}$ corresponds to a periodic orbit of $\mathcal{H}_{rot}$. To determine the attitude stability, we also introduce a deviation vector, 
\begin{equation}
\mathbf{w} = (\delta \theta, \delta \phi, \delta \psi, \delta P_\theta, \delta P_\phi, \delta P_\psi),	
\end{equation}
and the system of variational equations as, 
\begin{equation}
\dot{\mathbf{w}} = \mathcal{J} \mathbf{w},
\label{eq:vars}
\end{equation}
where $\mathcal{J}$ is the Jacobian of the flow. For a given trajectory $(r(t) \text{ and } u(t))$, the Hamiltonian equations for the rotation are solved along with the variational equations (Eq.~\eqref{eq:vars}) and the stability is determined from the fast Lyapunov indicator (FLI) defined as \citep{Froeschle1997,Skokos2010},
\begin{equation}
\textrm{FLI}(t) = \sup_t \log_{10} || \mathbf{w}(t) ||. 
\end{equation}

\subsection{Uncoupled natural frequencies}
\label{appendix: natural frequencies}
Let us assume a triaxial satellite, with moments of inertia ($A<B<C$), orbiting a primary on a Keplerian orbit. The natural rotational frequencies close to the synchronous state are approximated by the following expressions (see for example \cite{Fleig1970}):
\begin{align}
k &= 3 (1 + \frac{3}{2} e^2 + \frac{15}{8} e^4 ) + \mathcal{O}(e^6) \nonumber,\\
r_1 &= \frac{A}{C} \nonumber,\\
r_2 &= \frac{B}{C} \nonumber,\\
a &= r_1 r_2 \nonumber,\\
b & = k r_1^2 - 2 r_1 r_2 - (k-1) r_1 + r_2 -1, \nonumber\\
c & = (k+1) (1-r_1) (1-r_2), \nonumber\\
\omega_{\text{off},1} &= n \sqrt{\frac{-b - \sqrt{b^2-4 a c}}{2 a}}, \\
\omega_{\text{off},2} &= n \sqrt{\frac{-b + \sqrt{b^2-4 a c}}{2 a}}, \\
\omega_{\text{lib}} &= n \sqrt{k (r_2-r_1)},
\end{align}
where $n$ is the mean motion, $e$ is the eccentricity of the orbit, $\omega_{\text{lib}}$ is the planar libration frequency and $\omega_{\text{off},1}, \omega_{\text{off},2}$ are the two coupled off-plane frequencies related to the precession and nutation of the body. In the notation of this paper, we refer to $\omega_{\text{off},1}$ as $\omega^{uc}_{\text{prc}}$. 

\section{Initial conditions optimization scheme}
\label{appendix:optimization scheme}

Due to the small uncertainty in the observed binary orbit period, we prioritize having the simulated pre-impact Didymos system match the observed binary orbit period as close as possible. The non-Keplerian nature of the system requires the use of a numerical optimization scheme to derive the initial conditions. Our procedure is described below.

In order to construct the nominal pre-impact state for a given choice in the secondary's axial ratios, we make the simplifying assumption that the binary is in a dynamically relaxed state. This implies the following:

\begin{enumerate}
  \item The binary orbit is circular or nearly circular $(e\approx0.0)$.
  \item The mutual orbit pole is initially aligned with the primary's spin pole.
  \item The secondary's rotation is synchronous with its orbit (i.e., tidally locked), and its spin pole is aligned with the mutual orbit pole.
  \item The secondary's libration amplitude has been damped to a minimum.
\end{enumerate}

All of these assumptions are consistent (or at the very least not in disagreement) with current constraints of Didymos' orbit \citep{scheirich2009, naidu2020} and observations of other similar binary systems \citep{pravec2016}. We also assume that both bodies have a uniform mass distribution and the same bulk density, which is appropriate if they have the same origin or if the secondary was created through YORP spin-up driven mass loss of the primary. Upon future observations, if it is found that any of these assumptions are incorrect, this optimization scheme can be changed accordingly. 

Due to the non-spherical shapes and close proximity of the binary components, their motion is expected to be highly non-Keplerian. Therefore, estimating the binary mass with Kepler's 3$^{\text{rd}}$ Law is only accurate to ${\sim}1\%$, which is insufficient for reproducing the observed binary orbit period in simulations. Therefore, we implemented a simple optimization scheme to generate the initial conditions of the Didymos binary that best match the observed orbit period. The routine keeps the initial body positions, velocities, and spins fixed, while adjusting the bulk density (and therefore the total mass, moments of inertia, etc.)\ until the desired orbit period is achieved. The routine has the following steps:

\begin{enumerate}
  \item Set the primary's spin rate to its observed value and the secondary's spin rate equal to the \textit{observed} mean motion with both bodies' spin poles aligned.
  \item Set the separation between the mass centers equal to the observed semimajor axis.
  \item Set the relative velocity between the two bodies such that their instantaneous orbital angular velocity is equal to the observed mean motion. The velocity should be orthonormal to the radial separation vector and in the direction such that orbit pole and spin poles are aligned.
  \item Run a root-finding algorithm to determine the bulk density necessary to match the observed orbit period.
\end{enumerate}

\subsection*{Root-finding algorithm}
Once the relative positions, velocities, spins, and body orientations are set, we begin the optimization process. We use the secant method root-finding algorithm to find the bulk density that gives the correct orbit period. We are trying to find the root to the function,
\begin{equation}
   f(\rho) = P_{\text{sim}}(\rho) - P_{\text{obs}},
 \end{equation} 
 where $\rho$ is the system bulk density, $P_{\text{sim}}(\rho)$ is the simulated orbit period, and $P_{\text{obs}}$ is the observed orbit period. The secant method is a finite difference version of Newton's method, where each successive guess for the bulk density is given by,

 \begin{equation}
   \rho_{n} = \rho_{n-1} - f(\rho_{n-1})\frac{\rho_{n-1} - \rho_{n-2}}{f(\rho_{n-1}) - f(\rho_{n-2})}
 \end{equation}

Each step in the root-finding process requires running a full-2-body-problem code for several orbital periods in order to calculate $P_{\text{sim}}(\rho)$. However, the secant method converges quickly in this case, matching the observed orbit period to within $10^{-6}$ s in ${\sim}5$ iterations. Matching the orbit period to such high precision is not completely necessary, because the 3-sigma uncertainty on the measured orbit period is ${\sim}0.7$ s. However, the optimization process is not computationally expensive so it requires a negligible amount of time.

\begin{figure}
\centering
\includegraphics[width=.55\textwidth]{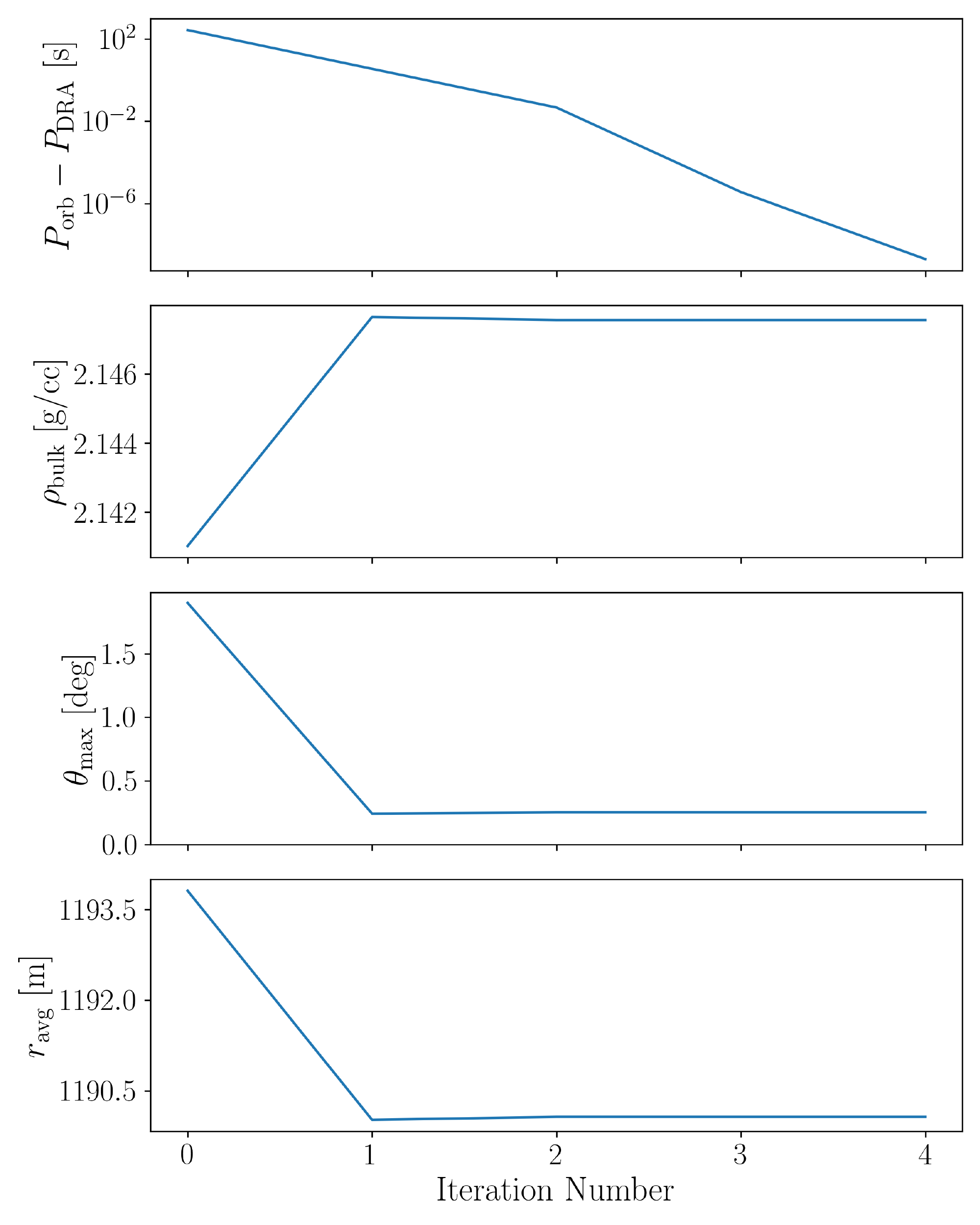}
\caption{\label{appendix: convergence} A plot of the difference between the simulated and observed orbit periods, bulk density, maximum libration amplitude, and average separation as a function of each iteration in the optimization process. After the first iteration, the scheme gets a good solution, but continues making minute adjustments to the bulk density (which aren't discernible on this plot) until the orbit period is matched to high precision.}
\end{figure}

An example of the scheme converging to a solution is shown in Fig.\ \ref{appendix: convergence}. The top panel shows how accurately the simulated orbit period matches the observed orbit period. The remaining three panels show the system bulk density, maximum libration amplitude, and average separation, respectively. The bulk density for the initial guess (Iteration \#0) is calculated using the 2nd-order equilibrium solution for a doubly synchronous binary (See Eq. 39 in \cite{scheeres2009}). Although the Didymos primary is not in synchronous rotation, this initial guess is a much better approximation than a Keplerian solution and is more than sufficient for a starting point.

\begin{figure}[t!]
    \centering
    \begin{subfigure}[t]{0.5\textwidth}
        \centering
        \includegraphics[width=\textwidth]{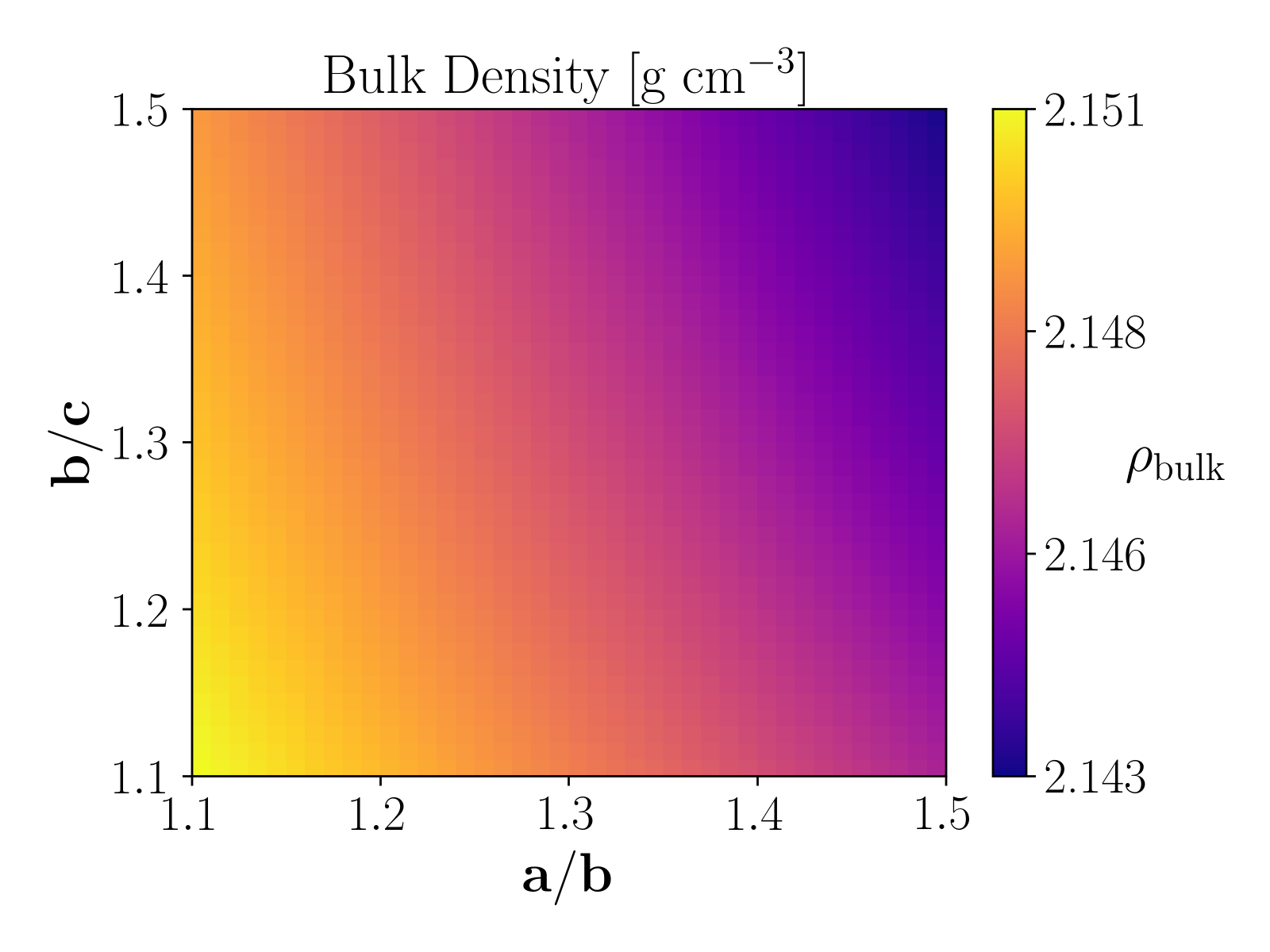}
        \caption{Optimized bulk density.\label{fig: rhoBulk_appendix}}
    \end{subfigure}%
    ~ 
    \begin{subfigure}[t]{0.5\textwidth}
        \centering
        \includegraphics[width=\textwidth]{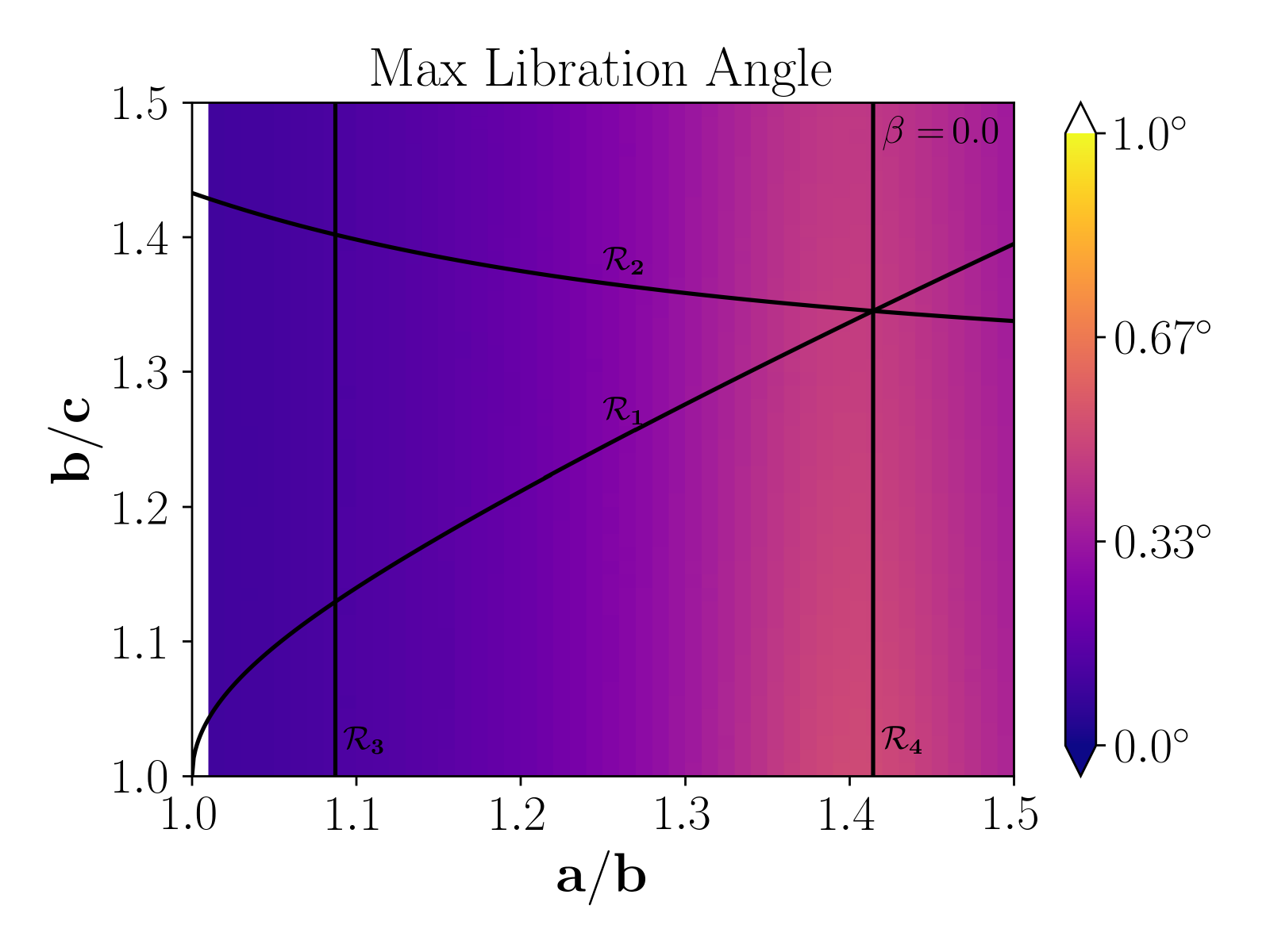}
        \caption{Resulting maximum libration amplitude.\label{fig: libAmp_appendix}}
    \end{subfigure}
  \caption{\label{appendix: optimization} (\ref{fig: rhoBulk_appendix}) The bulk density resulting from the optimization scheme. Each shape of the secondary has a slightly different bulk density (and total mass) but all simulations have the same orbit period. (\ref{fig: libAmp_appendix}) The resulting libration amplitude after each optimized initial condition is run for 1 year. The libration amplitude has been minimized, with the peak near $a/b\simeq1.4$ corresponding to the expected 1:1 resonance between the mean motion and free libration frequency ($\mathcal{R}_{4}$).}
\end{figure}

Despite conserving the same total volume in each simulation, changing the shape of the secondary changes the mass distribution and therefore the mutual potential energy, which affects the orbit period. This sensitivity to the initial conditions requires that the optimization scheme is run for each choice in the secondary's shape that we want to study. Fig.\ \ref{appendix: optimization} shows the optimized bulk density for each choice in the secondary's axial ratios along with the resulting maximum libration amplitude. These changes are all very small, so the system mass never differs by more than ${\sim}\frac{1}{2}\%$ between two simulations.

\end{document}